\documentclass[11pt, letterpaper, onecolumn, peerreview]{IEEEtran}

\usepackage{epsfig}
\usepackage{amssymb}
\usepackage[tbtags]{amsmath}
\usepackage{latexsym}
\usepackage{euscript}
\usepackage{graphics,eepic,epic,psfrag}

\usepackage[tbtags]{amsmath} 
\usepackage{amssymb}  
\usepackage{verbatim} 

\newcounter{actr}
{\begin{list}{(\alph{actr})}{\usecounter{actr}}}{\end{list}}

\newcounter{ictr}
{\begin{list}{(\roman{ictr})}{\usecounter{ictr}}}{\end{list}}

\newtheorem{theorem}{Theorem}
\newtheorem{lemma}{Lemma}
\newtheorem{definition}{Definition}
\newtheorem{corollary}{Corollary}

\newcommand{\qed}{\rule[0.1ex]{1.4ex}{1.6ex}}

\DeclareMathAlphabet{\mathbsf}{OT1}{cmss}{bx}{n}
\DeclareMathAlphabet{\mathssf}{OT1}{cmss}{m}{sl}

\DeclareSymbolFont{bsfletters}{OT1}{cmss}{bx}{n}
\DeclareSymbolFont{ssfletters}{OT1}{cmss}{m}{n}
\DeclareMathSymbol{\bsfGamma}{0}{bsfletters}{'000}
\DeclareMathSymbol{\ssfGamma}{0}{ssfletters}{'000}
\DeclareMathSymbol{\bsfDelta}{0}{bsfletters}{'001}
\DeclareMathSymbol{\ssfDelta}{0}{ssfletters}{'001}
\DeclareMathSymbol{\bsfTheta}{0}{bsfletters}{'002}
\DeclareMathSymbol{\ssfTheta}{0}{ssfletters}{'002}
\DeclareMathSymbol{\bsfLambda}{0}{bsfletters}{'003}
\DeclareMathSymbol{\ssfLambda}{0}{ssfletters}{'003}
\DeclareMathSymbol{\bsfXi}{0}{bsfletters}{'004}
\DeclareMathSymbol{\ssfXi}{0}{ssfletters}{'004}
\DeclareMathSymbol{\bsfPi}{0}{bsfletters}{'005}
\DeclareMathSymbol{\ssfPi}{0}{ssfletters}{'005}
\DeclareMathSymbol{\bsfSigma}{0}{bsfletters}{'006}
\DeclareMathSymbol{\ssfSigma}{0}{ssfletters}{'006}
\DeclareMathSymbol{\bsfUpsilon}{0}{bsfletters}{'007}
\DeclareMathSymbol{\ssfUpsilon}{0}{ssfletters}{'007}
\DeclareMathSymbol{\bsfPhi}{0}{bsfletters}{'010}
\DeclareMathSymbol{\ssfPhi}{0}{ssfletters}{'010}
\DeclareMathSymbol{\bsfPsi}{0}{bsfletters}{'011}
\DeclareMathSymbol{\ssfPsi}{0}{ssfletters}{'011}
\DeclareMathSymbol{\bsfOmega}{0}{bsfletters}{'012}
\DeclareMathSymbol{\ssfOmega}{0}{ssfletters}{'012}

\newcommand{\rvb}{{\mathssf{b}}}    

\newcommand{\rvs}{{\mathssf{s}}}    

\newcommand{\rvx}{{\mathssf{x}}}    

\newcommand{\svx}{x}            

\newcommand{\rvy}{{\mathssf{y}}}    

\newcommand{\svy}{y}

\newcommand{\rvxtil}{\tilde{\rvx}}

\newcommand{\rvxhat}{\hat{\rvx}}
\newcommand{\rvshat}{\hat{\rvs}}

\newcommand{\svxBBar}{\bar{\bar{\svx}}}

\newcommand{\xhat}{\widehat{\svx}}
\newcommand{\xtil}{\widetilde{\svx}}
\newcommand{\xBar}{\bar{\svx}}

\newcommand{\svyBBar}{\bar{\bar{\svy}}}

\newcommand{\delay}{\Delta}

\begin{document}
\title{The price of ignorance:\\The impact of side-information on delay for lossless source-coding}

\author{Cheng Chang and Anant Sahai\footnote{Both C.~Chang and
    A.~Sahai are with the Wireless Foundations Center within the
    Department of Electrical Engineering and Computer Science,
    University of California, Berkeley, CA 94720 (E-mails:
    \texttt{\{cchang,sahai\}@eecs.berkeley.edu}). This material was
    presented in part at the Information Workshop, Punta del Este,
    Uruguay March 2006 and the IEEE Int Symp Inform Theory, Seattle,
    USA, July 2006.}  }
\maketitle



\renewcommand{\baselinestretch}{1}

\begin{abstract}
  Inspired by the context of compressing encrypted sources, this paper
  considers the general tradeoff between rate, end-to-end delay, and
  probability of error for lossless source coding with
  side-information. The notion of end-to-end delay is made precise by
  considering a sequential setting in which source symbols are
  revealed in real time and need to be reconstructed at the decoder
  within a certain {\em fixed} latency requirement. Upper bounds are
  derived on the reliability functions with delay when
  side-information is known only to the decoder as well as when it is
  also known at the encoder.

  When the encoder is not ignorant of the side-information (including
  the trivial case when there is no side-information), it is possible
  to have substantially better tradeoffs between delay and probability
  of error at all rates. This shows that there is a fundamental {\em
    price of ignorance} in terms of end-to-end delay when the encoder
  is not aware of the side information. This effect is not visible if
  only fixed-block-length codes are considered. In this way,
  side-information in source-coding plays a role analogous to that of
  feedback in channel coding.

  While the theorems in this paper are asymptotic in terms of long
  delays and low probabilities of error, an example is used to show
  that the qualitative effects described here are significant even at
  short and moderate delays.
\end{abstract}
\renewcommand{\baselinestretch}{1}


\IEEEpeerreviewmaketitle

\section{Introduction}

There are two surprising classical results pertaining to encoder
``ignorance:'' Shannon's finding in \cite{ShannonZeroError} that the
capacity of a memoryless channel is unchanged if the encoder has
access to feedback and the Slepian-Wolf result in
\cite{SlepianWolf:73} that side-information at the encoder does not
reduce the data-rate required for lossless compression. When the rate
is not at the fundamental limit (capacity or conditional entropy), the
error probability converges to zero exponentially in the allowed
system delay --- with block-length serving as the traditional proxy
for delay in information theoretic studies.  Dobrushin in
\cite{DobrushinReliability} and Berlekamp in \cite{BerlekampThesis}
followed up on Shannon's result to show that feedback also does not
improve\footnote{The history of feedback and its impact on channel
  reliability is reviewed in detail in \cite{OurUpperBoundPaper}.} the
block-coding error exponent in the high-rate regime (close to
capacity) for symmetric channels.  Similarly, Gallager in
\cite{gallagerTech:76} and Csisz{\'{a}}r and K{\"{o}}rner in
\cite{csiszarkorner} showed that the block-coding error exponents for
lossless source-coding also do not improve with
encoder-side-information in the low rate regime (close to the
conditional entropy). These results seemed to confirm the overall
message that the advantages of encoder knowledge are limited to
possible encoder/decoder implementation complexity reductions, not to
anything more basic like rate or probability of error.

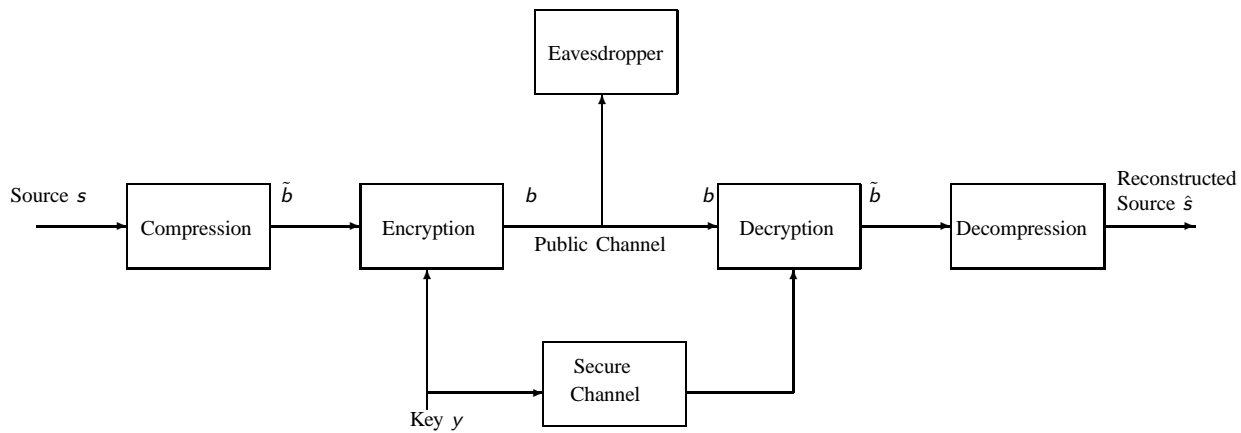
\begin{figure}[htbp]
\begin{center}
\setlength{\unitlength}{2750sp}%
\begingroup\makeatletter\ifx\SetFigFont\undefined%
\gdef\SetFigFont#1#2#3#4#5{%
  \reset@font\fontsize{#1}{#2pt}%
  \fontfamily{#3}\fontseries{#4}\fontshape{#5}%
  \selectfont}%
\fi\endgroup%
{\begin{picture}(11113,3816)(1000,-3790) \thinlines
{\put(1876,-2311){\framebox(1275,750){}}
}%
{ \put(3976,-2311){\framebox(1275,750){}}
}%
{ \put(7201,-2311){\framebox(1275,750){}}
}%
{ \put(5251,-1936){\vector( 1, 0){1950}}
}%
{ \put(8476,-1936){\vector( 1, 0){825}}
}%
{ \put(9301,-2311){\framebox(1375,750){}}
}%
{ \put(3151,-1936){\vector( 1, 0){825}}
}%
{ \put(1051,-1936){\vector( 1, 0){825}}
}%
{ \put(10676,-1936){\vector( 1, 0){825}}
}%
{ \put(5551,-736){\framebox(1275,750){}}
}%
{ \put(5626,-3736){\framebox(1275,750){}}
}%
{ \put(6151,-1936){\vector( 0, 1){1200}}
}%
{ \put(6901,-3436){\line( 1, 0){975}} \put(7876,-3436){\vector( 0,
1){1125}}
}%
{ \put(4576,-3586){\vector( 0, 1){1275}}
}%
{ \put(4576,-3436){\vector( 1, 0){1050}}
}%
\put(1926,-2011){\makebox(0,0)[lb]{\smash{{\SetFigFont{8}{14.4}{\rmdefault}{\mddefault}{\updefault}{ Compression}%
}}}}
\put(7326,-2011){\makebox(0,0)[lb]{\smash{{\SetFigFont{8}{14.4}{\rmdefault}{\mddefault}{\updefault}{ Decryption}%
}}}}
\put(4101,-2011){\makebox(0,0)[lb]{\smash{{\SetFigFont{8}{14.4}{\rmdefault}{\mddefault}{\updefault}{ Encryption }%
}}}}
\put(9276,-2011){\makebox(0,0)[lb]{\smash{{\SetFigFont{8}{14.4}{\rmdefault}{\mddefault}{\updefault}{ Decompression }%
}}}}
\put(5601,-436){\makebox(0,0)[lb]{\smash{{\SetFigFont{8}{14.4}{\rmdefault}{\mddefault}{\updefault}{ Eavesdropper}%
}}}}
\put(5831,-3256){\makebox(0,0)[lb]{\smash{{\SetFigFont{8}{14.4}{\rmdefault}{\mddefault}{\updefault}{ Secure}%
}}}}
\put(5801,-3506){\makebox(0,0)[lb]{\smash{{\SetFigFont{8}{14.4}{\rmdefault}{\mddefault}{\updefault}{ Channel}%
}}}}
\put(751,-1711){\makebox(0,0)[lb]{\smash{{\SetFigFont{8}{14.4}{\rmdefault}{\mddefault}{\updefault}{ Source  $\rvs$}%
}}}}
\put(10726,-1561){\makebox(0,0)[lb]{\smash{{\SetFigFont{8}{14.4}{\rmdefault}{\mddefault}{\updefault}{ Reconstructed}%
}}}}
\put(10726,-1786){\makebox(0,0)[lb]{\smash{{\SetFigFont{8}{14.4}{\rmdefault}{\mddefault}{\updefault}{ Source  $\hat{\rvs}$}%
}}}}
\put(4351,-3736){\makebox(0,0)[lb]{\smash{{\SetFigFont{8}{14.4}{\rmdefault}{\mddefault}{\updefault}{ Key  $\rvy$}%
}}}}
\put(5476,-2161){\makebox(0,0)[lb]{\smash{{\SetFigFont{8}{14.4}{\rmdefault}{\mddefault}{\updefault}{     Public Channel }%
}}}}

\put(3251,-1711){\makebox(0,0)[lb]{\smash{{\SetFigFont{8}{14.4}{\rmdefault}{\mddefault}{\updefault}{$\tilde{\rvb}$}%
}}}}

\put(5451,-1711){\makebox(0,0)[lb]{\smash{{\SetFigFont{8}{14.4}{\rmdefault}{\mddefault}{\updefault}{$\rvb$}%
}}}}

\put(7051,-1711){\makebox(0,0)[lb]{\smash{{\SetFigFont{8}{14.4}{\rmdefault}{\mddefault}{\updefault}{$\rvb$}%
}}}}

\put(8551,-1711){\makebox(0,0)[lb]{\smash{{\SetFigFont{8}{14.4}{\rmdefault}{\mddefault}{\updefault}{$\tilde{\rvb}$}%
}}}}

\end{picture}}
\caption[]{The traditional compression/encryption system for sources
with redundancy. (Figure adapted from \cite{JohnsonEncryption})}
\label{fig.encryption_setup1}
\end{center}
\end{figure}

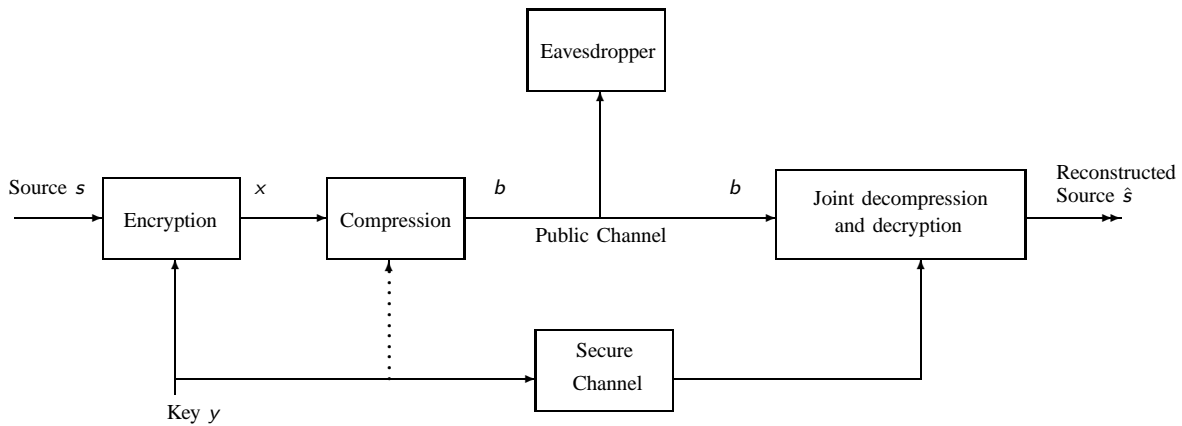
\begin{figure}[htbp]
\begin{center}
\setlength{\unitlength}{2650sp}%
\begingroup\makeatletter\ifx\SetFigFont\undefined%
\gdef\SetFigFont#1#2#3#4#5{%
  \reset@font\fontsize{#1}{#2pt}%
  \fontfamily{#3}\fontseries{#4}\fontshape{#5}%
  \selectfont}%
\fi\endgroup%
\begin{picture}(10825,3891)(1000,-3865)
\thinlines { \put(1876,-2311){\framebox(1275,750){}}
}%
{ \put(3976,-2311){\framebox(1275,750){}}
}%
{ \put(5251,-1936){\vector( 1, 0){2925}}
}%
{ \put(3151,-1936){\vector( 1, 0){825}}
}%
{ \put(1051,-1936){\vector( 1, 0){825}}
}%
{ \put(10576,-1936){\vector( 1, 0){825}}
}%
{ \put(7201,-3436){\line( 1, 0){2325}} \put(9526,-3436){\vector( 0,
1){1125}}
}%
{ \put(2551,-3436){\vector( 1, 0){3375}}
}%
{ \put(2551,-3586){\vector( 0, 1){1275}}
}%
{ \put(4551,-2406){\vector( 0, 1){75}}
}%

{ \multiput(4521,-3446)( 0, 100){11}{.}
}%

{ \put(8176,-2311){\framebox(2325,825){}}
}%
{ \put(10501,-1936){\vector( 1, 0){825}}
}%
{ \put(6526,-1936){\vector( 0, 1){1200}}
}%
{ \put(5851,-736){\framebox(1275,750){}}
}%
{ \put(5926,-3736){\framebox(1275,750){}}
}%
\put(10726,-1561){\makebox(0,0)[lb]{\smash{{\SetFigFont{8}{14.4}{\rmdefault}{\mddefault}{\updefault}{ Reconstructed}%
}}}}
\put(10726,-1786){\makebox(0,0)[lb]{\smash{{\SetFigFont{8}{14.4}{\rmdefault}{\mddefault}{\updefault}{       Source  $\hat{\rvs}$}%
}}}}
\put(2001,-2011){\makebox(0,0)[lb]{\smash{{\SetFigFont{8}{14.4}{\rmdefault}{\mddefault}{\updefault}{ Encryption}%
}}}}
\put(4026,-2011){\makebox(0,0)[lb]{\smash{{\SetFigFont{8}{14.4}{\rmdefault}{\mddefault}{\updefault}{ Compression}%
}}}}
\put(2401,-3811){\makebox(0,0)[lb]{\smash{{\SetFigFont{8}{14.4}{\rmdefault}{\mddefault}{\updefault}{ Key  $\rvy$}%
}}}}
\put(8451,-1811){\makebox(0,0)[lb]{\smash{{\SetFigFont{8}{14.4}{\rmdefault}{\mddefault}{\updefault}{ Joint decompression }%
}}}}
\put(8651,-2061){\makebox(0,0)[lb]{\smash{{\SetFigFont{8}{14.4}{\rmdefault}{\mddefault}{\updefault}{       and decryption}%
}}}}
\put(926,-1711){\makebox(0,0)[lb]{\smash{{\SetFigFont{8}{14.4}{\rmdefault}{\mddefault}{\updefault}{ Source  $\rvs$}%
}}}}
\put(5851,-2161){\makebox(0,0)[lb]{\smash{{\SetFigFont{8}{14.4}{\rmdefault}{\mddefault}{\updefault}{     Public Channel }%
}}}}
\put(5901,-436){\makebox(0,0)[lb]{\smash{{\SetFigFont{8}{14.4}{\rmdefault}{\mddefault}{\updefault}{ Eavesdropper}%
}}}}
\put(6301,-3236){\makebox(0,0)[lb]{\smash{{\SetFigFont{8}{14.4}{\rmdefault}{\mddefault}{\updefault}{Secure  }%
}}}}
\put(6201,-3536){\makebox(0,0)[lb]{\smash{{\SetFigFont{8}{14.4}{\rmdefault}{\mddefault}{\updefault}{ Channel}%
}}}}

\put(3226,-1711){\makebox(0,0)[lb]{\smash{{\SetFigFont{8}{14.4}{\rmdefault}{\mddefault}{\updefault}{   $\rvx$}%
}}}}

\put(5526,-1711){\makebox(0,0)[lb]{\smash{{\SetFigFont{8}{14.4}{\rmdefault}{\mddefault}{\updefault}{$\rvb$}%
}}}}

\put(7726,-1711){\makebox(0,0)[lb]{\smash{{\SetFigFont{8}{14.4}{\rmdefault}{\mddefault}{\updefault}{$\rvb$}%
}}}}
\end{picture}%
\caption{The novel compression/encryption system in which a message
  is first encrypted and then compressed by the ``ignorant'' encoder.
  (Figure adapted from \cite{JohnsonEncryption}) }
\label{fig.encryption_setup2}
\end{center}
\end{figure}

Once low complexity channel codes were developed that did not need
feedback, mathematical and operational duality (See
e.g.~\cite{coverChiang:02,pradhanChouRamchandran:03}) enabled
corresponding advances in low complexity distributed source codes.
These codes then enabled radical new architectures for media coding in
which the complexity could be shifted from the encoder to the decoder
\cite{pradhanRamchandran:03, girod:05}. Even more provocatively,
\cite{JohnsonEncryption} introduced a new architecture for
information-theoretic secure communication illustrated as a shift from
Figure~\ref{fig.encryption_setup1} to
Figure~\ref{fig.encryption_setup2}. By viewing Shannon's one-time-pad
from \cite{ShannonEncryption} as virtual side information, Johnson,
{\em et al} in \cite{JohnsonEncryption} showed that despite being
marginally white and uniform, encrypted data could be compressed just
as effectively by a system that does not have access to the key, as
long as decoding takes place jointly with decryption. However, all of
this work followed the traditional fixed-block-length perspective on
source and channel coding.

Recently, it has become clear that the behavior of fixed-block-length
codes and fixed-delay codes can be quite different in contexts where
the message to be communicated is revealed to the encoder gradually as
time progresses rather than being known all at once. In our entire
discussion, the assumption is that information arises as a stream
generated in real time at the source (e.g.~voice, video, or sensor
measurements) and it is useful to the destination in finely grained
increments (e.g.~a few milliseconds of voice, a single video frame,
etc.). The encoded bitstream is also assumed to be transported at a
steady rate. The acceptable end-to-end delay is determined by the
application and can often be much larger than the natural granularity
of the information being communicated (e.g.~voice may tolerate a delay
of hundreds of milliseconds despite being useful in increments of a
few milliseconds). The end-to-end delay perspective here is common in
the networking community. This is different from cases in which
information arises in large bursts with each burst needing to be
received by the destination before the next burst even becomes
available at the source.

\cite{OurUpperBoundPaper} shows that unlike the block channel coding
reliability functions, the reliability function with respect to {\em
  fixed} end-to-end delay can in fact improve dramatically with
feedback for essentially all DMCs at high rates.\footnote{It had long
  been known that the reliability function with respect to {\em
    average} block-length can improve \cite{burnashev}, but there was
  a mistaken assertion by Pinsker in \cite{PinskerNoFeedback} that the
  fixed-delay exponents do not improve with feedback.} The asymptotic
factor reduction in end-to-end delay enabled by feedback approaches
infinity as the message rate approaches capacity for generic DMCs. In
addition, the nature of the dominant error events changes. Consider
time relative to when a message symbol enters the encoder. Without
feedback, errors are usually caused by {\em future} channel
atypicality. When feedback is present, it is a {\em combination of
  past and future} atypicality that causes errors.

The results in \cite{OurUpperBoundPaper} give a precise interpretation
to the channel-coding half of Shannon's intriguingly prophetic comment
at the close of \cite{ShannonLossy}:
\begin{quotation}
  ``[the duality of source and channel coding] can be pursued further
  and is related to a duality between past and future and the notions
  of control and knowledge. Thus we may have knowledge of the past and
  cannot control it; we may control the future but have no knowledge
  of it.''
\end{quotation}

One of the side benefits of this paper is to make Shannon's comment
similarly precise on the source coding side. Rather than worrying
about what the appropriate granularity of information should be, the
formal problem is specified at the individual source symbol level. If
a symbol is not delivered correctly by its deadline, it is considered
to be erroneous. The upper and lower bounds of this paper turn out to
not depend on the choice of information granularity, only on the fact
that the granularity is much finer than the tolerable end-to-end
delay.

Here, we show that when decoder side-information is also available at
the encoder, the dominant error event involves only the {\em past}
atypicality of the source. This gives an upper bound on the
fixed-delay error exponent that is the lossless source-coding
counterpart to the ``uncertainty focusing bound'' given in
\cite{OurUpperBoundPaper} for channel coding with feedback. This bound
is also shown to be asymptotically achievable at all rates. When
side-information is present only at the decoder,
\cite{StreamingSlepianWolf} showed that the much slower random-coding
error exponent is attainable with end-to-end delay.  Here, an upper
bound is given on the error exponent that matches the random-coding
bound from \cite{StreamingSlepianWolf} at low rates for appropriately
symmetric cases --- like the case of compressing encrypted data from
\cite{JohnsonEncryption}. This shows that there is a fundamental price
of encoder ignorance that must be paid in terms of required end-to-end
delay.

Section~\ref{sec:notationresults} fixes notation, gives the problem
setup, and states the main results of this paper after reviewing the
relevant classical results. Section~\ref{sec:numericexample} evaluates
a specific numerical example to show the penalties of encoder
ignorance. It also demonstrates how the delay penalty continues to be
substantial even in the non-asymptotic regime of short end-to-end
delays and moderately small probability of error requirements.
Section~\ref{sec:source_ei} gives the proof for the fixed delay
reliability function when both encoder and decoder have access to
side-information. Section~\ref{sec:source_si} proves the
upper-bound on the fixed-delay reliability function when the encoder
is ignorant of the side-information and the appendices show that it is
tight for the symmetric case. Finally, Section~\ref{sec:conclusion}
gives some concluding remarks by pointing out the parallels between
the source and channel coding stories.

\section{Notation, problem setup and main results}
\label{sec:notationresults}

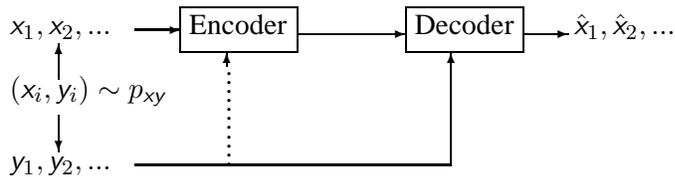
\begin{figure}[htbp]
\setlength{\unitlength}{0.6mm}
\begin{picture}(100,40)(-60,0)

\put(40,30){\line(1,0){26}} \put(40,40){\line(1,0){26}}
\put(40,30){\line(0,1){10}} \put(66, 30){\line(0,1){10}}
\put(42,34){Encoder }

\put(90,30){\line(1,0){26}} \put(90,40){\line(1,0){26}}
\put(90,30){\line(0,1){10}} \put(116, 30){\line(0,1){10}}
\put(92,34){Decoder }

\put(116, 34){\vector(1,0){10}} \put(66, 34){\vector(1,0){24}}

\put(30, 5){\line(1,0){70}}
 \put(100, 5){\vector(0,1){25}}
 \put(50.3,27){\vector(0,1){3}}

  \multiput(50,5)( 0, 2){11}{.}

\put(125,34){ $\hat{\rvx}_1,\hat{\rvx}_2,...$}

\put(30, 35){\vector(1,0){10}}

\put(0, 34){ ${\rvx}_1,{\rvx}_2,...$} \put(0,4){
${\rvy}_1,{\rvy}_2,...$}

\put(0, 20){ $(\rvx_i,\rvy_i)\sim p_{\rvx\rvy}$}

\put(13,24){\vector(0,1){8}} \put(13,16){\vector(0,-1){8}}
 \end{picture}
     \caption{Lossless source coding with encoder/decoder side-information.}
     \label{fig.encoder_side_information}
 \end{figure}

In this paper, all sources are iid random processes from finite
alphabets where the finite alphabets are identified with the first few
non-negative integers. $\rvx$ and $\rvy$ are random variables taking
values in $\mathcal{X}$ and $\mathcal{Y}$, with $x$ and $y$ used to
denote realizations of the random variables.  Without loss of
generality, assume that $\forall x\in {\mathcal X}, \forall y \in
{\mathcal Y}$, the marginals $p_\rvx(x)>0$ and $p_\rvy(y)>0$. The
basic problem formulation is illustrated in
Figure~\ref{fig.encoder_side_information} for the cases with or
without encoder access to the side-information.

 The goal is to losslessly communicate the source $\rvx$, drawn from a
 joint distribution $p_{\rvx\rvy}$ on $\rvx,\rvy$, over a fixed rate
 bit-pipe. The decoder is always assumed to have access to the
 side-information $\rvy$, and it may or may not be available to the
 encoder as well.

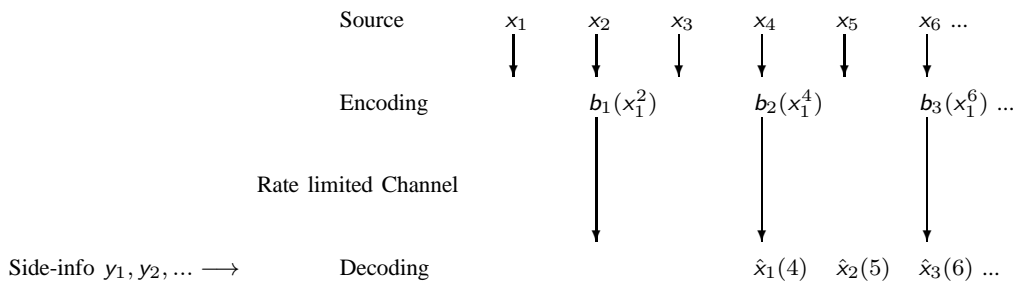
\begin{figure}[htbp]
\setlength{\unitlength}{1mm}
\begin{center}
\scalebox{1.1}{\begin{picture}(140,40)(-40,-5)

\scriptsize
 \put(20, 30){$\rvx_1$} \put(30, 30){$\rvx_2$} \put(40,
30){$\rvx_3$} \put(50, 30){$\rvx_4$} \put(60, 30){$\rvx_5$} \put(70,
30){$\rvx_6$ ...}

  \put(30, 20){$\rvb_1(\rvx_1^2)$}
  \put(50, 20){$\rvb_2(\rvx_1^4)$}
  \put(70, 20){$\rvb_3(\rvx_1^6)$ ...}

\put(50, 0){$\rvxhat_1(4)$} \put(60,  0){$\rvxhat_2(5)$} \put(70,
 0){$\rvxhat_3(6)$ ...}

 \put(0,  0){Decoding}

\put(0, 20){Encoding}

\put(0, 30){Source}

\put(-10, 10){Rate limited Channel}

\put(-40,  0){Side-info $\rvy_1,\rvy_2,...\longrightarrow$}
\multiput(21,29)(10,0){6}{\vector(0,-1){5}}
\multiput(31,19)(20,0){3}{\vector(0,-1){15}}
    \end{picture}}
    \caption{Time line for fixed-delay source coding with
      decoder side-information: rate $R=\frac{1}{2}$, delay $\delay=3$. }
    \label{fig:time_line}
\end{center}
\end{figure}

Rather than being known entirely in advance, the source symbols enter
the encoder in a real-time fashion. (Illustrated in
Figure~\ref{fig:time_line}) For convenience, time is counted in terms
of source symbols: we assume that the source $\mathcal{S}$ generates a
pair of source symbols $(\rvx,\rvy)$ per second from the finite
alphabet $\mathcal{X}\times\mathcal{Y}$. The $j$'th source symbol
$\rvx_j$ is not known at the encoder until time $j$ and similarly for
$\rvy_j$ at the decoder (and possibly encoder).  Rate $R$ operation
means that the encoder sends $1$ binary bit to the decoder every
$\frac{1}{R}$ seconds. Throughout the paper the focus is on cases with
$H_{\rvx|\rvy} < R < \log_2|\mathcal{X}|$, since the lossless coding
problem becomes trivial outside of that range.

\begin{definition}
  A rate $R$ encoder $\mathcal{E}$ is a sequence of maps
  $\{\mathcal{E}_j\},j=1,2,\ldots$. The outputs of $\mathcal{E}_j$ are
  the bits that are communicated from time $j-1$ to $j$. When the
  encoder does not have access to the decoder side-information:
\begin{eqnarray}
&&\mathcal{E}_j: \mathcal{X}^{j} \longrightarrow \{0,1\}^{\lfloor jR\rfloor-\lfloor(j-1)R\rfloor}\nonumber\\
&&\mathcal{E}_j(x_1^j)=b_{\lfloor (j-1)R \rfloor+1}^{\lfloor jR
\rfloor}\nonumber
\end{eqnarray}
When the encoder does have access to the decoder side-information:
\begin{eqnarray}
&&\mathcal{E}_j: \mathcal{X}^{j}\times\mathcal{Y}^{j} \longrightarrow \{0,1\}^{\lfloor jR\rfloor-\lfloor(j-1)R\rfloor}\nonumber\\
&&\mathcal{E}_j(x_1^j,y_1^j)=b_{\lfloor (j-1)R \rfloor+1}^{\lfloor jR \rfloor}\nonumber
\end{eqnarray}
\end{definition}

\begin{definition}
  A fixed delay $\delay$ decoder $\mathcal{D}^\delay$ is a sequence of
  maps $\{\mathcal{D}^\delay_j\},j=1,2,\ldots$. The input to
  $\mathcal{D}^\delay_j$ are the all the bits emitted by the encoder
  until time $j+\delay$ as well as the side-information
  $\rvy_1^{j+\delay}$. The output is an estimate $\widehat{\rvx}_j$ for
  the source symbol $\rvx_j$. 

  Alternatively, a family of decoders indexed by different delays can
  be considered together. For these, the output is a list
  $\widehat{\rvx}(j) = (\widehat{\rvx}_1(j),\ldots,\widehat{\rvx}_j(j))$. 
\begin{eqnarray*}
\mathcal{D}^{\delay}_j:& \{0,1 \}^{\lfloor jR \rfloor}\times\mathcal{Y}^j \longrightarrow
  \mathcal{X} \\
\mathcal{D}^{\delay}_j(b_{1}^{\lfloor jR
\rfloor},y_1^j)& =\widehat{x}_{j-\delay}(j)
\end{eqnarray*}
where $\widehat{x}_{j-\delay}(j)$ is the estimate of $\rvx_{j-\delay}$
at time $j$ and thus has an end-to-end delay of $\delay$ seconds.
\end{definition}

The problem of lossless source-coding is considered by examining the
asymptotic tradeoff between delay and the probability of symbol error:

\begin{definition} \label{def:delayconstrained_ee} A family
  (indexed by delay $\delay$) of rate $R$ sequential source codes
  $\{({\mathcal E}^\delay, {\mathcal D}^{\delay}\}$ achieves
  fixed-delay reliability $E(R)$ if for all $\epsilon > 0$, there
  exists $K < \infty$, s.t. $\forall i, \delay >0$
$$ \Pr(\rvx_i \neq
\widehat{\rvx}_i(i+\delay)) \leq K 2^{-\delay(E(R)-\epsilon)}$$
when encoder ${\mathcal E}^\delay$ is used to do the encoding of the
source and ${\mathcal D}^{\delay}\}$ is the decoder used to recover
$\widehat{\rvx}$.
\end{definition}

It is important to see that all source positions $i$ require equal
protection in terms of probability of error, but the probability of
error can never be made uniform over the source realizations
themselves since it is the source that is the main source of
randomness in the problem! 

\subsection{Review of block source coding with side information }

Before stating the new results, it is useful to review the classical
fixed-block-length coding results. In fixed-block-length coding, the
encoder has access to $\rvx_1^n$ all at once (as well as $\rvy_1^n$ if
it has access to the side-information) and produces $nR$ bits all at
once.  These bits go to the block decoder along with the
side-information $\rvy_1^n$ and the decoder then produces estimates
$\widehat{\rvx}_1^n$ all at once. While the usual error probability
considered is the block error probability $\Pr(\rvx_1^n\neq
\hat{\rvx}_1^n)=\Pr(\rvx_1^n\neq
\mathcal{D}_n(\mathcal{E}_n(\rvx_1^n)))$, there is no difference
between the symbol error probability and the block-error probability
on an exponential scale.

The relevant error exponents $E(R)$ are considered in the limit of
large block-lengths, rather than end-to-end delays.
$E(R)$ is achievable if $\exists$ a family of
$\{(\mathcal{E}_n,\mathcal{D}_n)\}$, s.t.
\begin{equation}
\lim_{n\rightarrow \infty}-\frac{ 1}{n}\log_2 \Pr(\rvx_1^n\neq
\widehat{\rvx}_1^n)=E(R)
\end{equation}

The relevant results of \cite{csiszarkorner, gallagerTech:76} are summarized
into the following theorem.
\begin{theorem}\label{THM.SI_BLOCK} If the block-encoder does not have access
  to the side-information, the best possible block-error exponent is
  sandwiched between two bounds: $E_{si, b}^{l}(R)\leq E_{si,b}(R)\leq
  E_{si,b}^{u}(R)$ where
\begin{eqnarray}
E_{si,b}^{l}(R)
&=&\min_{q_{\rvx\rvy}}\{D(q_{\rvx\rvy}\|p_{\rvx\rvy})+\max\{0, R-
H(q_{\rvx|\rvy})\}\} \label{eqn:blocklowerbound}\\
&=&\sup_{0\leq \rho\leq 1 } \rho R - E_0(\rho) \label{eqn:blocklowerboundrho}\\
E_{si,b}^{u}(R) &=&\min_{q_{\rvx\rvy}: H(q_{\rvx|\rvy})\geq
R} D(q_{\rvx\rvy}\|p_{\rvx\rvy}) \label{eqn:blockupperbound}\\
&=&\sup_{0\leq \rho} \rho R - E_0(\rho) \label{eqn:blockupperboundrho}
\end{eqnarray}
where
\begin{equation} \label{eqn:gallagerfunction}
E_0(\rho) = \log_2\sum_{y
}(\sum_{x}p_{\rvx\rvy}(x,y)^{\frac{1}{1+\rho}})^{(1+\rho)}
\end{equation}
is the Gallager function for the source with side-information.

The lower-bound corresponds to the performance of random-binning
with MAP decoding. The upper and lower bounds agree for rates close
to $H(p_{\rvx|\rvy})$, specifically $R \leq \frac{\partial
  E_0(\rho)}{\partial \rho}|_{\rho=1}$. 

If the encoder also has access to the side-information, then
$E_{si,b}^{u}(R)$ is the true error exponent at all rates since it
can be achieved by simply encoding the conditional type of $x_1^n$
given $y_1^n$ and then encoding the index of the true realization
within that conditional type.

\end{theorem}

If there is no side-information, then $y=0$ and the problem behaves
like the case of side-information known at the encoder.
(\ref{eqn:blockupperbound}) recovers the simple point-to-point
fixed-block-length error exponent for lossless source coding.  The
resulting random and non-random error exponents are:
\begin{eqnarray}
E^r_{s,b}(R,p_\rvx)
&=&\min_{q_{\rvx}}\{D(q_{\rvx}\|p_{\rvx})+\max\{0, R-
H(q_{\rvx})\}\} \label{eqn:blocklowerbound_singlesource}\\
E_{s,b}(R,p_\rvx) &=&\min_{q_{\rvx}: H(q_{\rvx})\geq
R} D(q_{\rvx}\|p_{\rvx}). \label{eqn:blockupperbound_singlesource}
\label{eqn:blockupperboundrho_singlesource}
\end{eqnarray}
The Gallager function (\ref{eqn:gallagerfunction}) in
(\ref{eqn:blockupperboundrho}) and (\ref{eqn:blocklowerboundrho})
also simplifies to 
\begin{equation} \label{eqn:gallagerfunction_singlesource}
E_0(\rho) = (1+\rho)\log_2(\sum_{x}p_{\rvx}(x)^{\frac{1}{1+\rho}}).
\end{equation}

\subsection{Main results}

\cite{StreamingSlepianWolf} shows that the random coding bound
$E^l_{si,b}(R)$ is achievable with respect to end-to-end delay even
without the encoder having access to the side-information. Thus, the
factor of two increase in delay caused by using a fixed-block-length
code in a real-time context is unnecessary.
\cite{StreamingSlepianWolf} uses a randomized sequential binning
strategy with either MAP decoding or a universal decoding scheme that
works for any iid source. \cite{HariSequentialPaper} shows that the
same asymptotic tradeoff with delay is achievable using a more
computationally friendly stack-based decoding algorithm if the
underlying joint distribution is known. However, it turns out that the
end-to-end delay performance can be much better if the encoder has
access to the side-information.

\begin{theorem}\label{THM_EI} For fixed rate $R$ lossless source-coding of an
  iid source with side-information present at both the receiver and
  encoder, the asymptotic error exponent $E_{ei}(R)$ with fixed
  end-to-end delay is given by the source uncertainty-focusing bound:
\begin{equation} \label{eqn:focusingsource}
 E_{ei}(R) = \inf_{\alpha>0}
 \frac{1}{\alpha}E_{si,b}^u((\alpha+1) R)
\end{equation}
where $E_{si,b}^u$ is defined in (\ref{eqn:blockupperbound}) and
(\ref{eqn:blockupperboundrho}). The source uncertainty-focusing bound
can also be expressed parametrically in terms of the Gallager function
$E_0(\rho)$ from (\ref{eqn:gallagerfunction}):
\begin{eqnarray}
 E_{ei}(R) & = & E_0(\rho) \nonumber\\
  R    & = & \frac{E_0(\rho)}{\rho}\label{eqn:parameterization_of_Eei}
\end{eqnarray}

This bound generically approaches $R = H(\rvx|\rvy)$ at strictly
positive slope of $2 H(\rvx|\rvy)/ \frac{\partial^2 E_0(0)}{\partial
  \eta^2}$. When $\frac{\partial^2 E_0(0)}{\partial \eta^2} = 0$, the
fixed-delay reliability function jumps discontinuously from zero to
infinity. 

Furthermore, this bound is asymptotically achievable by using
universal fixed-to-variable block codes whose resulting data bits are
smoothed to fixed-rate $R$ through a FIFO queue with an infinite
buffer size. This code is universal over iid sources as well as
end-to-end delays that are sufficiently long (the block-length for the
code is much smaller than the asymptotically large end-to-end delay
constraint). 
\end{theorem}
\vspace{0.1in}

\begin{theorem}\label{THM_UPPER_SI} For fixed rate $R$ lossless
  source-coding of an iid source with side-information {\em only} at
  the receiver, the asymptotic error exponent $E_{si}(R)$ with fixed
  end-to-end delay must satisfy $E_{si}(R) \leq E_{si}^{u}(R)$, where
\begin{eqnarray}
E_{si}^{u}(R)= &\min\big\{&\inf_{q_{\rvx\rvy}, \alpha\geq 1:
H(q_{\rvx|\rvy})>(1+\alpha)R}
\frac{1}{\alpha}D(q_{\rvx\rvy}\|p_{\rvx\rvy}),\nonumber\\
&& \inf_{q_{\rvx\rvy}, 1\geq \alpha\geq 0:
H(q_{\rvx|\rvy})>(1+\alpha)R} \frac{1-\alpha}{\alpha}D(q_{\rvx}\|p_{\rvx})
+D(q_{\rvx\rvy}\|p_{\rvx\rvy}) \big\} \label{eqn:upperboundstreaming}
\end{eqnarray}
\end{theorem}
\vspace{0.1in}

\begin{figure}[htbp]
\setlength{\unitlength}{3947sp}%
\begingroup\makeatletter\ifx\SetFigFont\undefined%
\gdef\SetFigFont#1#2#3#4#5{%
  \reset@font\fontsize{#1}{#2pt}%
  \fontfamily{#3}\fontseries{#4}\fontshape{#5}%
  \selectfont}%
\fi\endgroup%
\begin{center}
\begin{picture}(4488,2010)(1576,-3811)
\thinlines { \put(2400,-2161){\vector( 1, 0){3200}}
}%
{ \put(2400,-2211){\vector( 3,-1){3200}}
}%
{ \put(2400,-3361){\vector( 1, 0){3200}}
}%
{ \put(2400,-3311){\vector( 3, 1){3200}}
}%
\put(1576,-2761){\makebox(0,0)[lb]{\smash{{\SetFigFont{12}{14.4}{\rmdefault}{\mddefault}{\updefault}{$\rvy$}%
}}}}
\put(5926,-2836){\makebox(0,0)[lb]{\smash{{\SetFigFont{12}{14.4}{\rmdefault}{\mddefault}{\updefault}{$\rvx$}%
}}}}
\put(3700,-1936){\makebox(0,0)[lb]{\smash{{\SetFigFont{12}{14.4}{\rmdefault}{\mddefault}{\updefault}{$1-\epsilon$}%
}}}}
\put(3700,-3811){\makebox(0,0)[lb]{\smash{{\SetFigFont{12}{14.4}{\rmdefault}{\mddefault}{\updefault}{$1-\epsilon$}%
}}}}
\put(2800,-3136){\makebox(0,0)[lb]{\smash{{\SetFigFont{12}{14.4}{\rmdefault}{\mddefault}{\updefault}{$\epsilon$}%
}}}}
\put(2800,-2511){\makebox(0,0)[lb]{\smash{{\SetFigFont{12}{14.4}{\rmdefault}{\mddefault}{\updefault}{$\epsilon$}%
}}}}

\put(2200,-2200){\makebox(0,0)[lb]{\smash{{\SetFigFont{12}{14.4}{\rmdefault}{\mddefault}{\updefault}{$0$}%
}}}}

\put(2200,-3400){\makebox(0,0)[lb]{\smash{{\SetFigFont{12}{14.4}{\rmdefault}{\mddefault}{\updefault}{$1$}%
}}}}

\put(5650,-2200){\makebox(0,0)[lb]{\smash{{\SetFigFont{12}{14.4}{\rmdefault}{\mddefault}{\updefault}{$0$}%
}}}}

\put(5650,-3400){\makebox(0,0)[lb]{\smash{{\SetFigFont{12}{14.4}{\rmdefault}{\mddefault}{\updefault}{$1$}%
}}}}

\end{picture}%
\caption{A joint distribution on $\rvx,\rvy$ that comes from a
  discrete memoryless channel connecting the two together and where
  the $\rvy$ is uniform and independent of the channel. }
    \label{fig:DMC_MODEL}
\end{center}
\end{figure}

%
%
%
%
%
For symmetric cases (such as those depicted in
Figure~\ref{fig:DMC_MODEL}), we have the following corollary:

\begin{corollary} \label{corollary_SI_upper_symmetric}
Consider iid $(\rvx, \rvy) \sim p_{\rvx\rvy}$ such that the
side-information $\rvy$ is uniform on ${\cal Y}$ and $\rvx=\rvy \oplus
\rvs$, where $\rvs \sim p_\rvs$ is independent of $\rvy$. Then the
asymptotic error exponent $E_{si}(R)$ with fixed delay must satisfy
$E_{si}(R) \leq E_{si,b}^{u}(R)=E_{s,b}(R,p_\rvs)$ from
(\ref{eqn:blockupperbound}) and
(\ref{eqn:blockupperbound_singlesource}).
\end{corollary}

Since \cite{StreamingSlepianWolf} shows that $E_{si,b}^{u}(R)$ is
achievable at low rates, Corollary~\ref{corollary_SI_upper_symmetric}
is tight there.


%
%

\section{Application and numeric example}\label{sec:numericexample}

While the above results are general, they can be applied to the
specific context of the \cite{JohnsonEncryption} approach of
compressing encrypted data. The general problem is depicted in
Figure~\ref{fig:time_line_encryption_with_delay} in terms of joint
encryption and compression. The goal is to communicate from end-to-end
using a reliable fixed-rate bit-pipe in such a way that:
\begin{itemize}
 \item The required rate of the bit-pipe is low.
 \item The probability of error is low for each source symbol.
 \item The end-to-end delay is small.
 \item Nothing is revealed to an eavesdropper that has access to the
       bitstream.
\end{itemize}
The idea is to find a good tradeoff among the first three while
preserving the fourth. To support these goals, assume access to an
infinite supply of common-randomness shared among the encoder and
decoder that is not available to the eavesdropper. This can be used as
a secret key. We are not concerned here with the size of the secret
key.

This section evaluates the fixed-delay performance for both the
compression-first and encryption-first systems as a way of showing the
delay price of the encoder's ignorance of the side-information in the
encryption-first approach. Nonasymptotic behavior is explored using a
specific code for short short values of end-to-end delay to verify
that the price of ignorance also hits when delays are small.

\subsection{Encryption/compression  of streaming data: asymptotic results}\label{sec:encryptionresults}

The main results of this paper can be used to evaluate two candidate
architectures: the traditional compression-first approach depicted in
Figure~\ref{fig.encryption_setup1} and the novel encryption-first
approach proposed in \cite{JohnsonEncryption} and depicted in
Figure~\ref{fig.encryption_setup2}.

\begin{figure}[htbp]
\setlength{\unitlength}{1mm}
\begin{center}
\scalebox{1.1}{\begin{picture}(140,30)(-30,0)

\scriptsize
 \put(20, 30){$\rvs_1$} \put(30, 30){$\rvs_2$} \put(40,
30){$\rvs_3$} \put(50, 30){$\rvs_4$} \put(60, 30){$\rvs_5$} \put(70,
30){$\rvs_6$ ...}

  \put(30, 20){$\rvb_1(\rvs_1^2)$}
  \put(50, 20){$\rvb_2(\rvs_1^4)$}
  \put(70, 20){$\rvb_3(\rvs_1^6)$ ...}

\put(-20, 20){Compression/Encryption}

\put(0, 30){Source}

  \multiput(21,29)(10,0){6}{\vector(0,-1){5}}

\put(50,  0){$\rvshat_1(4)$} \put(60,  0){$\rvshat_2(5)$} \put(70,
 0){$\rvshat_3(6)$ ...}

\put(-10, 10){Rate limited channel}
   \put(-20, 0){Decompression/Decryption}

\multiput(31,19)(20,0){3}{\vector(0,-1){15}}
    \end{picture}}
    \caption{ Joint encryption and compression  of streaming data with
      a fixed end-to-end delay constraint. Here the rate $R=\frac{1}{2}$
      bits per source symbol and delay $\delay=3$.}
    \label{fig:time_line_encryption_with_delay}
\end{center}
\end{figure}

\subsubsection{Compress and then encrypt}
The traditional compression-first approach is immediately covered by
Theorem~\ref{THM_EI} since the lack of
side-information as far as compression is concerned can be modeled by
having trivial side-information $Y=0$ and $\rvx = \rvs$. In that case,
the relevant error exponent with end-to-end delay is given by
(\ref{eqn:parameterization_of_Eei}). The secret key can simply be used
to XOR the rate $R$ bitstream with a one-time-pad.

\begin{figure}[htbp]
\setlength{\unitlength}{1mm}
\begin{center}
\scalebox{1.1}{\begin{picture}(140,60)(-30,-30)

\scriptsize
 \put(20, 30){$\rvs_1$} \put(30, 30){$\rvs_2$} \put(40,
30){$\rvs_3$} \put(50, 30){$\rvs_4$} \put(60, 30){$\rvs_5$} \put(70,
30){$\rvs_6$ ...}

  \put(30, 20){$\tilde\rvb_1(\rvs_1^2)$}
  \put(50, 20){$\tilde\rvb_2(\rvs_1^4)$}
  \put(70, 20){$\tilde\rvb_3(\rvs_1^6)$ ...}

 \put(30, 10){$\rvb_1=\rvy_1\oplus \tilde\rvb_1$}
  \put(50, 10){$\rvb_2=\rvy_2\oplus \tilde\rvb_2$}
  \put(70, 10){$\rvb_3=\rvy_3\oplus \tilde\rvb_3$ ...}

 \put(30, -10){$\tilde\rvb_1=\rvy_1\oplus \rvb_1$}
  \put(50, -10){$\tilde\rvb_2=\rvy_2\oplus\rvb_2$}
  \put(70, -10){$\tilde\rvb_3=\rvy_3\oplus \rvb_3$ ...}

\put(50, -20){$\rvshat_1(4)$} \put(60, -20){$\rvshat_2(5)$} \put(70,
-20){$\rvshat_3(6)$ ...}

  \put(-30, 10){Secret key $\rvy_1, \rvy_2,...\longrightarrow$   Encryption}

 \put(-30, -10){Secret key $\rvy_1, \rvy_2,...\longrightarrow$   Decryption}

\put(0, 20){Compression}

\put(0, 30){Source} \put(-25,  0){Rate limited public channel
  }

\put(0, -20){Decompression}

  \multiput(21,29)(10,0){6}{\vector(0,-1){5}}
\multiput(31,19)(20,0){3}{\vector(0,-1){5}}
\multiput(31,9)(20,0){3}{\vector(0,-1){15}}
\multiput(31,-11)(20,0){3}{\vector(0,-1){5}}
    \end{picture}}
    \caption{The traditional approach of compression followed by
      encryption, for fixed-delay encoding at rate $R=\frac{1}{2}$,
      delay $\delay=3$. } 
    \label{fig:time_line_encrytion_compressed_with_delay}
\end{center}
\end{figure}
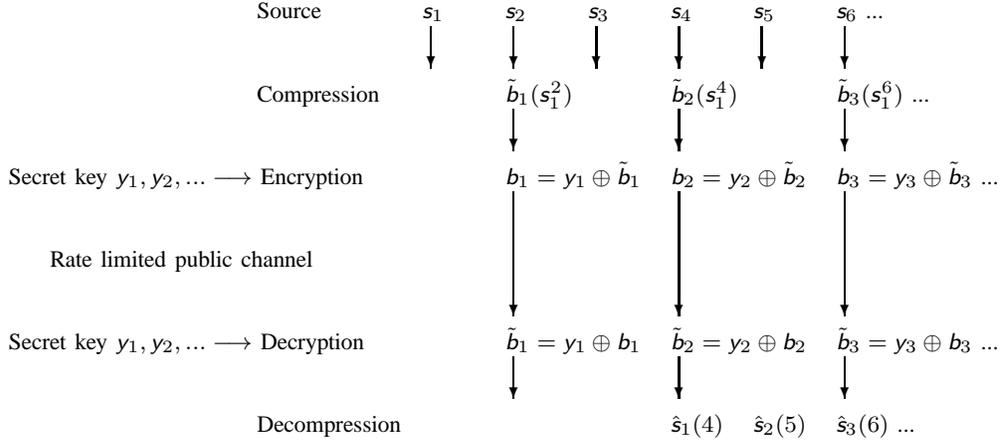

\subsubsection{Encrypt and then compress}
For the new approach of \cite{JohnsonEncryption}, the secret key is
used at a rate of $\log_2 |{\mathcal S}|$ bits per source-symbol to
generate iid uniform virtual side-information random variables $\rvy$
on the alphabet ${\mathcal Y} = {\mathcal X} = {\mathcal S}$. The
virtual source is generated by $\rvx = \rvs \oplus \rvy$ where the $+$
operation is interpreted in the Abelian group modulo $|{\mathcal S}|$.
It is clear from \cite{ShannonEncryption} that the mutual information
$I(\rvs_1^n;\rvx_1^n)=0$ for all $n$ and since there is a Markov chain $\rvs
\rightarrow \rvx \rightarrow \rvb$ to the encoded data bits, the
eavesdropper learns nothing about the source symbols. Given knowledge
of the secret key $\rvy$, decoding $\rvx$ correctly is equivalent to
decoding $\rvs$ correctly. Thus, the conditional entropy $H(\rvx|\rvy)
= H(\rvx \ominus \rvy| \rvy) = H(\rvs|\rvy) = H(\rvs)$ so nothing is lost in
terms of compressibility.

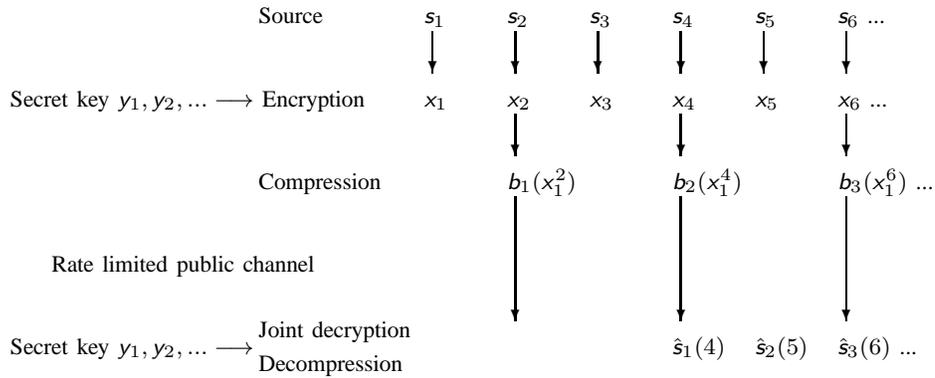
\begin{figure}[htbp]
\setlength{\unitlength}{1mm}
\begin{center}
\scalebox{1.1}{\begin{picture}(140,50)(-40,-20)

\scriptsize
 \put(20, 30){$\rvs_1$} \put(30, 30){$\rvs_2$} \put(40,
30){$\rvs_3$} \put(50, 30){$\rvs_4$} \put(60, 30){$\rvs_5$} \put(70,
30){$\rvs_6$ ...}

  \put(20, 20){$ \rvx_1  $}
  \put(30, 20){$ \rvx_2 $}
  \put(40, 20){$ \rvx_3 $ }
  \put(50, 20){$ \rvx_4 $}
  \put(60, 20){$ \rvx_5 $}
  \put(70, 20){$ \rvx_6 $ ...}

  \put(30, 10){$ \rvb_1(\rvx_1^2)$}
  \put(50, 10){$ \rvb_2(\rvx_1^4)$}
  \put(70, 10){$ \rvb_3(\rvx_1^6)$ ...}


\put(50, -10){$\rvshat_1(4)$} \put(60, -10){$\rvshat_2(5)$} \put(70,
-10){$\rvshat_3(6)$ ...}

  \put(-30, 20){Secret key $\rvy_1, \rvy_2,...\longrightarrow$   Encryption}

 \put(-30, -10){Secret key $\rvy_1, \rvy_2,...\longrightarrow$}

\put(0, 10){Compression}

\put(0, 30){Source} \put(-25,  0){Rate limited public channel
  }
\put(0, -8){Joint decryption} \put(0, -12){Decompression}

  \multiput(21,29)(10,0){6}{\vector(0,-1){5}}
\multiput(31,19)(20,0){3}{\vector(0,-1){5}}
\multiput(31,9)(20,0){3}{\vector(0,-1){15}}
    \end{picture}}
    \caption{Fixed-delay compression of already encrypted streaming data at
      rate $R=\frac{1}{2}$ and delay
      $\delay=3$.}    \label{fig:time_line_compressing_encrypted_with_delay} 
\end{center}
\end{figure}

Meanwhile, the marginal distributions for both the encrypted data
$\rvx$ and the secret key $\rvy$ are uniform. Since the conditions for
Corollary~\ref{corollary_SI_upper_symmetric} hold, the upper bound on
the error exponent with delay is $E_{si,b}^{u}(R)$ from
(\ref{eqn:blockupperbound}) and (\ref{eqn:blockupperboundrho}).
\cite{StreamingSlepianWolf} guarantees that a sequential random
binning strategy can achieve the exponent in
(\ref{eqn:blocklowerbound}) and (\ref{eqn:blocklowerboundrho}).

This means that nothing higher than the fixed-block-length error
exponent for source coding can be achieved with respect to end-to-end
delay if the encryption-first architecture is adopted with the
requirement that nothing about the true source be revealed to the
compressor. As the next section illustrates by example, there is a
severe delay price to requiring the encoder to be ignorant of the
source.

In practical terms, this means that if both the end-to-end delay and
acceptable probability of symbol error are constrained by the
application, then the approach of encryption followed by compression
can end up requiring higher-rate bit-pipes.

%

\subsection{Numeric example including nonasymptotic results}
Consider a simple source $\rvs$ with alphabet size $3$,
$\mathcal{S}=\{A, B, C \}$ and distribution 
\begin{eqnarray}
p_\rvs(A)= a \ \ \ p_\rvs(B)= \frac{1-a}{2} \ \ \  p_\rvs(C)=
\frac{1-a}{2} \nonumber
\end{eqnarray}
where $a=0.65$ for the plots and numeric comparisons.

\subsubsection{Asymptotic error exponents}

\begin{figure}[htbp]
\begin{center}
 \includegraphics[width=100mm]{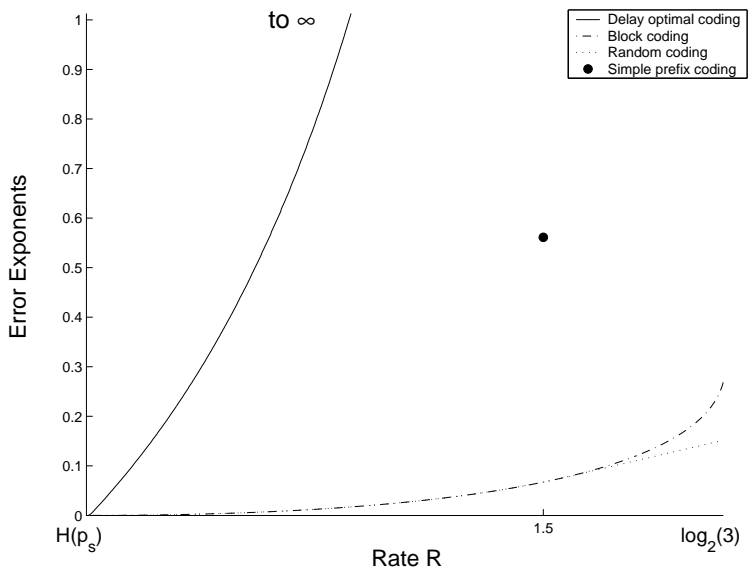}
\end{center}
\caption{Different source coding error exponents: fixed-delay error
  exponent $E_{s}(R)$ with encoder side-information,
  fixed-block-length error exponent $E_{s,b}(R, p_\rvs)$, and the
  random coding bound $E^r_{s,b}(R,p_\rvs)$. The fixed-block-length
  bound also bounds the fixed-delay case without encoder
  side-information since the example here is symmetric.} \label{fig.plots_source_coding}
\end{figure}

The different error exponents for fixed-block-length and fixed-delay
source coding predict the asymptotic performance of different source
coding systems when the end-to-end delay is long. We plot the source
uncertainty-focusing bound $E_{ei}(R)$, the fixed-block-length error
exponent $E_{s,b}(R,p_\rvs)$ and the random coding bound
$E^r_{s,b}(R,p_\rvs)$ in Figure~\ref{fig.plots_source_coding}. For
this source, the random coding and fixed-block-length error exponents
are the same for $R\leq\frac{\partial E_0(\rho)}{\partial
  \rho}|_{\rho=1}=1.509$.  Theorem~\ref{THM.SI_BLOCK} and
Theorem~\ref{THM_EI} reveal that these error exponents govern the
asymptotic performance of fixed-delay systems with and without encoder
side-information.


\begin{figure}[htbp]
\begin{center}
 \includegraphics[width=100mm]{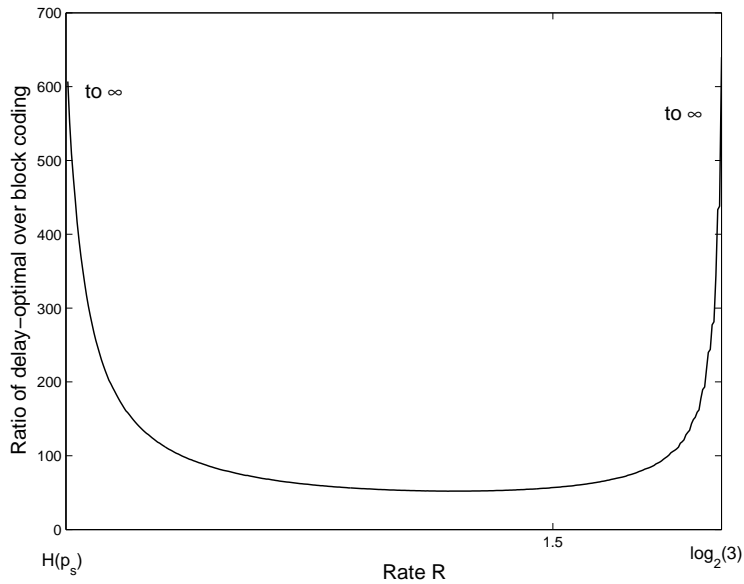}
 \caption{Ratio of the fixed-delay error exponent with encoder
   side-information $E_{ei}(R)$ over the 
   fixed-block-length error exponent $ E_{s,b}(R, p_\rvs)$. This
   reflects the asymptotic factor increase in end-to-end delay
   required to compensate for the encoder being ignorant of the
   side-information available at the decoder.}
\label{fig.plots_source_coding_ratio}
\end{center}
\end{figure}

Figure~\ref{fig.plots_source_coding_ratio} plots the ratio of the
source uncertainty-focusing bound over the fixed-block-length error
exponent. The ratio tells asymptotically how many times longer the
delay must be for the system built around an encoder that does not
have access to the side-information. The smallest ratio is around $52$
at a rate around $1.45$. 

\subsubsection{Non-asymptotic results}\label{sec:stream_prefix}

The price of ignorance is so high, that even non-optimal codes with
encoder side-information can outperform optimal codes without it. This
section uses a very simple fixed-delay coding scheme using a
prefix-free fixed-to-variable code\cite{CoverThomas} instead of the
asymptotically optimal universal code described in
Theorem~\ref{THM_EI}. The input block-length is two, and the encoder
uses the side-information to recover $\rvs$ before encoding it as: 
\begin{eqnarray*}
&&AA\rightarrow 0 \\
&&AB\rightarrow 1000\ \ AC\rightarrow 1001\ \ BA\rightarrow 1010\ \
BB\rightarrow 1011\ \ \\
&&BC\rightarrow 1100\ \ CA\rightarrow 1101\ \ CB\rightarrow 1110\ \
CC\rightarrow 1111\ \
\end{eqnarray*}

\begin{figure}[htbp]
\begin{center}
 \includegraphics[width=160mm]{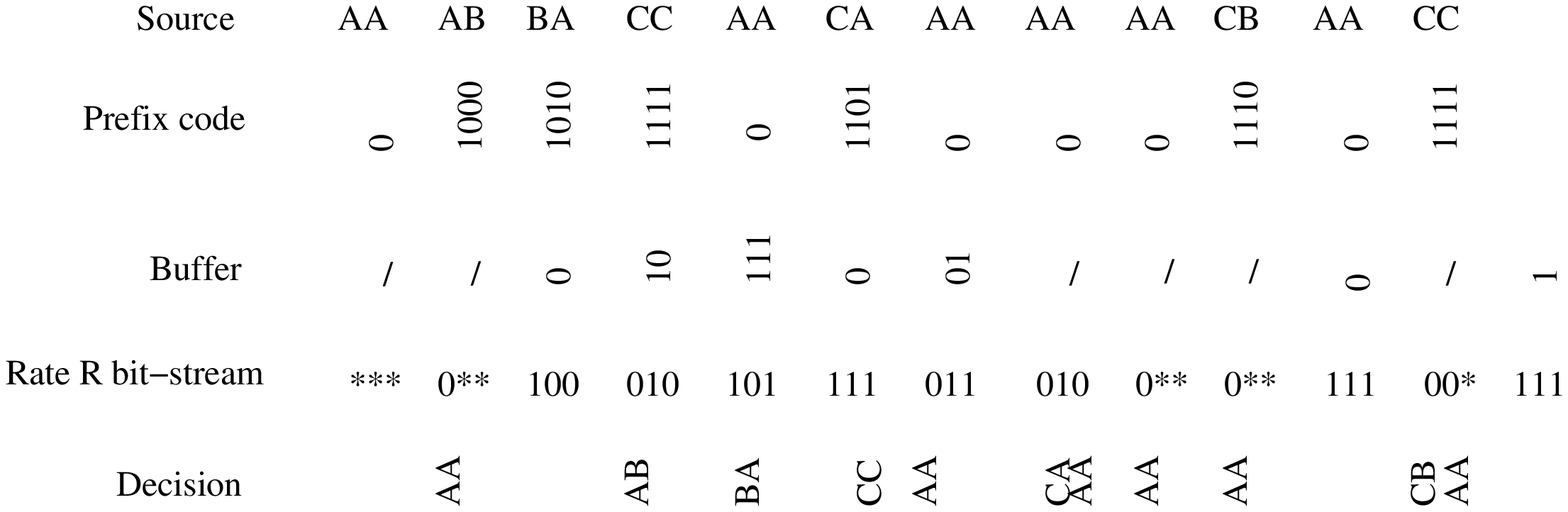}
\caption[]{ Suboptimal prefix coding system in action. / indicates empty queue,
* indicates meaningless filler bits.} \label{fig.prefix_stream}
\end{center}
\end{figure}

For ease of analysis, the system is run at $R=\frac{3}{2} <
\frac{\partial E_0(\rho)}{\partial \rho}|_{\rho=1}=1.509$. This means
that the source generates $1$ symbol per second and $3$ bits are sent
through the error-free bit-pipe every $2$ seconds. The variable-rate
of the code is smoothed through a FIFO queue with an infinite buffer
in a manner similar to the buffer-overflow problem studied in
\cite{JelinekBuffer, Merhav1991}. The entire coding system is
illustrated in Figure~\ref{fig.prefix_stream}.

It is convenient to examine time in increments of two seconds. The
length of the codeword generated is either $1$ or $4$. The buffer is
drained out by $3$ bits per $2$ seconds. Let $L_k$ be the number of
bits in the buffer as at time $2k$. Every two seconds, the number of
bits $L_{k}$ in the buffer either goes down by $2$ if $s_{2k-1},
s_{2k}= AA$ or goes up by $1$ if $s_{2k-1} s_{2k}\neq AA$. If the
queue is empty, the encoder can send arbitrary bits through the
bit-pipe without causing confusion at the decoder because the decoder
knows that the source only generates $1$ source symbol per second and
that it is caught up.

\begin{figure}[htbp]
\begin{center}
 \includegraphics[width=160mm]{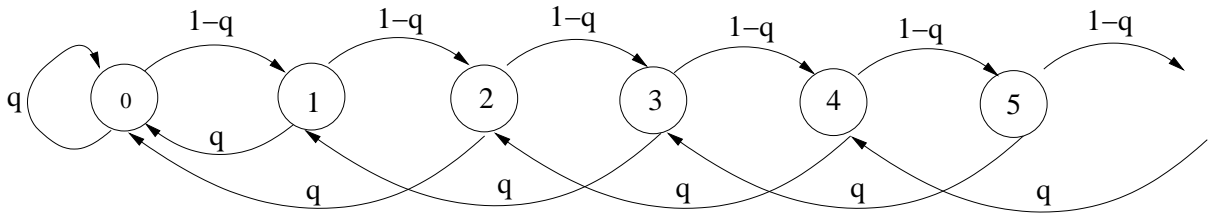}
 \caption[]{Transition graph of a reflecting random walk $L_k$ for
   queue length given the specified prefix-free code and the ternary
   source distribution $\{a, \frac{1-a}{2}, \frac{1-a}{2}\}$ and fixed
   rate $\frac{3}{2}$ bits per source symbol. $q=a^2$
   denotes the probability that $L_k$ decrements by 2.}
\label{fig.Markov_transition}
\end{center}
\end{figure}

Clearly $L_{k}, k=1,2,...$ forms a Markov chain with following
transition matrix: $L_{k}=L_{k-1}+1$ with probability $1-a^2$,
$L_{k}=L_{k-1}-2$ with probability $a^2$. The state transition graph
is illustrated in Figure~\ref{fig.Markov_transition}. For this Markov
chain, the stationary distribution can be readily
calculated\footnote{The polynomial corresponding to the recurrence
  relation for the stationary distribution has three roots. One of
  them is $1$ and the other is unstable since it has magnitude larger
  than $1$. That leaves only one possibility for the stationary
  distribution.} \cite{Durrett}.
\begin{eqnarray}\label{eqn:stationary}
\pi_{k}=Z \big(\frac{-1+\sqrt{1+\frac{4(1-q)}{q}}}{2}\big)^k
\end{eqnarray}
Where $q=a^2$ and $Z=1- \frac{-1+\sqrt{1+\frac{4(1-q)}{q}}}{2}$ is the
normalization constant. For this example $Z=0.228$. Notice that
$\pi_k$ is geometric and the stationary distribution exists as long as
$4\frac{1-q}{q}<8$ or equivalently $q>\frac{1}{3}$. In this example,
$a=0.65$ and thus $q=a^2=0.4225 > \frac{1}{3}$, so the stationary
distribution $\pi_k$ exists.

Assume $\delay$ is odd for convenience. For the above simple coding
system, a decoding error can only happen if at time $2k-1+\delay$, at
least one bit of the codeword describing $\rvs_{2k-1}, \rvs_{2k}$ is
still in the queue. Since the queue is FIFO, this implies that there
were too many bits awaiting transmission at time $2k$ itself --- ie
that the number of bits $L_{k}$ in the buffer at time $2k$, is larger
than
$$\lfloor\frac{3}{2}(\delay-1)\rfloor - l(\rvs_{2k-1}, \rvs_{2k})$$
where $l(\rvs_{2k-1}, \rvs_{2k})$ is the length of the codeword for
$\rvs_{2k-1}, \rvs_{2k}$. $l$ is $1$ with probability $q=a^2$ and $4$
with probability $1-q=1-a^2$. Notice that the length of the codeword
for $\rvs_{2k-1}, \rvs_{2k}$ is independent of $L_k$ since the
source symbols are iid. This gives the following upper bound on the
error probability of decoding with delay $\delay$ when the system is
in steady state\footnote{If the system is initialized to start in the
  zero state, then this bound remains valid since the system
  approaches steady state from below.}:
\begin{eqnarray*}
\Pr(\hat\rvs_{2k}(2k-1+\delay)\neq \rvs_{2k})&=&\nonumber\\
\Pr(\hat\rvs_{2k-1}(2k-1+\delay)\neq \rvs_{2k-1})&\leq& \Pr(l(\rvs_{2k-1}, \rvs_{2k})=1)\Pr(L_k> \lfloor\frac{3}{2}(\delay-1)\rfloor - 1) \\
&&  \Pr(l(\rvs_{2k-1}, \rvs_{2k})=4)\Pr(L_k>
\lfloor\frac{3}{2}(\delay-1)\rfloor - 4) \\
&=& q \sum_{j=\lfloor\frac{3}{2}(\delay-1)\rfloor }^\infty \pi_j+
(1-q) \sum_{j=\lfloor\frac{3}{2}(\delay-1)\rfloor -3}^\infty
\pi_j\\
&=& G
\big(\frac{-1+\sqrt{1+\frac{4(1-q)}{q}}}{2}\big)^{\lfloor\frac{3}{2}(\delay-1)\rfloor
-3}
\end{eqnarray*}
where $G$ is the normalization constant
\begin{eqnarray}
G= Z\big(q\sum_{j=0}^{2}(\frac{-1+\sqrt{1+\frac{4(1-q)}{q}}}{2})^j
+(1-q)\big).\nonumber
\end{eqnarray}
For this example, $G=0.360$. Thus, the fixed-delay error 
exponent for this coding system is
$$\frac{3}{2}\log_2 \big(\frac{-1+\sqrt{1+\frac{4(1-q)}{q}}}{2}\big).$$

\begin{figure}[htbp]
\begin{center}
 \includegraphics[width=100mm]{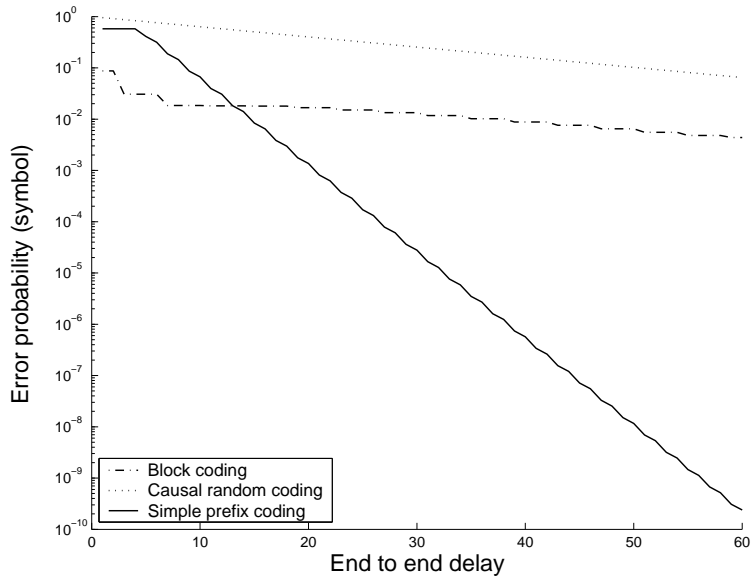}
\caption{Error probability vs delay (non-asymptotic results)
  illustrating the price of encoder ignorance.}
\label{fig.non_asymptotic}
\end{center}
\end{figure}

Figure~\ref{fig.non_asymptotic} compares three different coding
schemes in the non-asymptotic regime of short delays and moderate
probabilities of error at rate $\frac{3}{2}$. As shown in
Figure~\ref{fig.plots_source_coding}, at this rate the random coding
error exponent $E^r_{s,b}(R,p_\rvs)$ is the same as the
fixed-block-length error exponent $E_{s,b}(R, p_\rvs)$. The block
coding curve plotted is for an optimal coding scheme in which the
encoder first buffers up $\frac{\delay}{2}$ symbols, encodes them into
a length $\frac{\delay}{2}R$-length binary sequence and uses the next
$\frac{\delay}{2}$ seconds to transmit the message. This coding scheme
gives an error exponent $\frac{E_{s,b}(R, p_\rvs)}{2}$ with delay in
the limit of long delays.

The slope of these curves in Figure~\ref{fig.non_asymptotic} indicates
the error exponent governing how fast the error probability goes to
zero with delay. Although smaller than the delay optimal error
exponent $E_{s}(R)$, this simple coding strategy has a much higher
fixed-delay error exponent than both sequential random coding and
optimal \textit{simplex} block coding. A simple calculation reveals
that in order to get a $10^{-6}$ symbol error probability, the delay
requirement for our simple scheme is $\sim 40$, for causal random
coding is around $\sim 303$, and for optimal block coding is around
\footnote{We ran a linear regression on the data
  $y_\delay=\log_{10}P_e(\delay)$, $x_\delay=\delay$ as shown in
  Figure~\ref{fig.plots_source_coding} from $\delay=80$ to
  $\delay=100$ to extrapolate the $\delay$, s.t. $\log_{10}
  P_e(\delay)=-6$.} $\sim 374$. Thus, the price of encoder ignorance
is very significant even in the non-asymptotic regime and
fixed-block-length codes are very suboptimal from an end-to-end delay
point of view.

\section{Encoders with side-Information}\label{sec:source_ei}
The goal of this section is to prove Theorem~\ref{THM_EI} directly. 

\subsection{Achievability}
The achievability of $E_{ei}(R)$ is shown using a simple
fixed-to-variable\footnote{Fixed-to-variable was chosen for ease of
  analysis. It is likely that variable-to-fixed and
  variable-to-variable length codes can also be used as the basis for
  an optimal fixed-delay source coding system.}  length universal code
that has its output rate smoothed through a FIFO queue. Because the
end-to-end delay experienced by a symbol is dominated by the time
spent waiting in the queue, and the queue is drained at a
deterministic rate, the end-to-end delay experienced by a symbol is
essentially proportional to the length of the queue when that symbol
arrives. Thus on the achievability side, Theorem~\ref{THM_EI} can be
viewed as a corollary to results on the buffer-overflow exponent for
fixed-to-variable length codes. The buffer-overflow exponent was first
derived in \cite{JelinekBuffer} for cases without any side-information
at all. Here, we simply state the coding strategy used and leave the
detailed analysis for Appendix~\ref{app:achievabilityqueue}.

The strategy only depends on the size of the source alphabets
$|\mathcal X|, |\mathcal Y|$, not on the distribution of the 
source.

\begin{figure}[htbp]
\begin{center}
 \includegraphics[width=160mm]{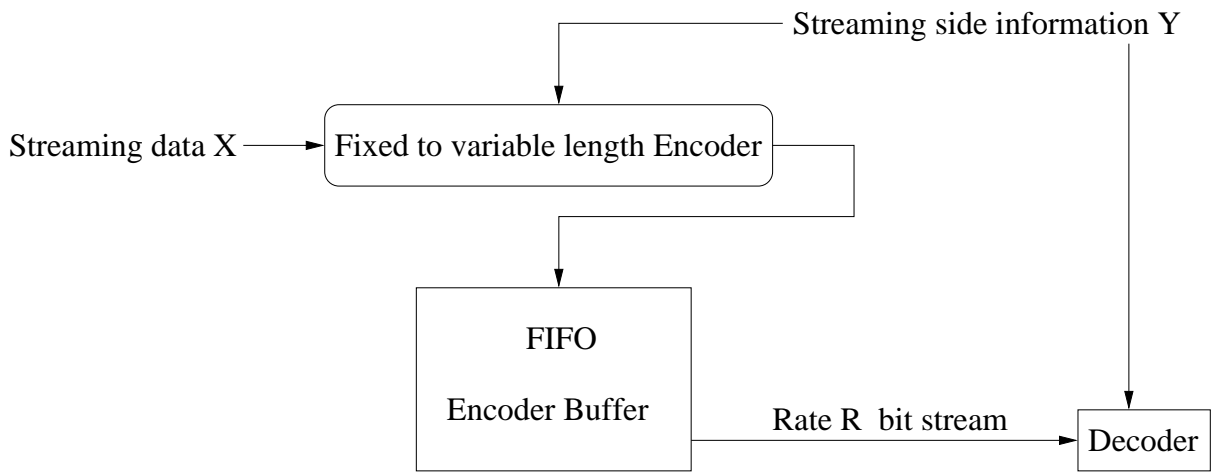}
\caption{ A universal fixed-delay lossless source coding system built
  around a fixed-to-variable block-length code.  }
\label{fig.universal_delay_coding_ei}
\end{center}
\end{figure}

First, a finite block-length $N$ is chosen that is much smaller than
the asymptotically large target end-to-end delay $\delay$. For a
discrete memoryless source $\rvx$, side information $\rvy$ and large
block-length $N$, an optimal fixed-to-variable code is given in
\cite{csiszarkorner} and consists of three stages: 
\begin{enumerate}
 \item Start with a $1$.

 \item Describe the joint type of the block $\vec{\rvx}_i$ (the
   $i$'th block of length $N$) and $\vec{\rvy}_i$. This costs at most
   a fixed $1 + |{\cal X}||{\cal Y}| \log_2 N$ bits per block.

 \item Describe which particular realization has occurred for
   $\vec{\rvx}_i$ by using a variable $NH(\vec{\rvx}_i|\vec{\rvy}_i)$
   bits where $H(\vec{\svx}_i|\vec{\svy}_i)$ is the empirical
  conditional entropy of sequence $\vec{\svx}_i$ given
  $\vec{\svy}_i$. 
\end{enumerate}

This code is obviously prefix-free. When the queue is empty, the
fixed-rate $R$ encoder can send a $0$ without introducing any
ambiguity. The total end-to-end delay experienced by any individual
source-symbol is then upper-bounded by $N$ (how long it must wait to
be assembled into a block) plus $\frac{1}{R}$ times the length of the
queue once it has been encoded.

Write $l(\vec{\rvx}_i,\vec{\rvy}_i)$ as the random total length of the
codeword for $\vec{\rvx}_i, \vec{\rvy}_i$. Then
\begin{eqnarray}
N H(\vec{\rvx}_i| \vec{\rvy}_i) \leq 
l(\vec{\rvx}_i, \vec{\rvy}_i)    =
N(H(\vec{\rvx}_i| \vec{\rvy}_i) + \epsilon_N)\label{eqn:lower_upperboundonl(s)_ei}
\end{eqnarray}
where $\epsilon_N \leq \frac{2 + |{\cal X}||{\cal Y}| \log_2
(N+1)}{N} $ goes to $0$ as $N$ gets large.  

Because the source is iid, the lengths of the blocks are also
iid. Each one has a length whose distribution can be bounded using
Theorem~\ref{THM.SI_BLOCK}. From there, there are two paths to show
the desired result. One path uses Corollary~6.1 of
\cite{OurUpperBoundPaper} and for that, all that is required is a
lemma parallel to Lemma~7.1 of \cite{OurUpperBoundPaper} asserting
that the length of the block has a distribution upperbounded by a
constant plus a geometric random variable. Such a bound easily follows
from the (\ref{eqn:blockupperboundrho}) formulation for the
block-reliability function. We take a second approach proceeding
directly using standard large deviations techniques. The following
lemma bounds the probability of atypical source behavior for the sum
of lengths.

\begin{lemma} \label{lemma:source_atyp_Source_ei} for all
  $\epsilon>0$, there exists a block length $N$ large enough so that
  there exists $K<\infty$ such that for all $n > 0$ and all $H(\rvx|\rvy) < r <
  \log_2|\mathcal{X}|$
\begin{eqnarray}
 \Pr(\sum_{i=1}^n l(\vec{\rvx}_i, \vec{\rvy}_i)
> nN r ) \leq K
2^{-n N(E^{u}_{si,b}(r)-\epsilon)}.\label{eqn:source_atypicality_ei}
\end{eqnarray}
\end{lemma}
\proof:  See Appendix~\ref{app:achievabilityqueue}.
\vspace{0.1in}

At time $(t+\delay)N$, the decoder \textit{cannot} decode
$\vec{\rvx}_t$ with $0$ error probability iff the binary strings
describing $\vec{\rvx}_t$ are \textit{not} all out of the buffer
yet. Since the encoding buffer is FIFO, this means that the number of
outgoing bits from some time $t_1$ to $(t+\delay)N$ is less than the
number of the bits in the buffer at time $t_1$ plus the number of
incoming bits from time $t_1$ to time $tN$. Suppose the buffer were
last empty at time $t_1 = tN-nN$ where $0 \leq n \leq t$. Given this,
a decoding error could occur only if $ \sum_{i=0}^{n-1}
l(\vec{\rvx}_{t-i},\vec{\rvy}_{t-i}) > (n+\delay)NR $.

Denote the longest code length by  $l_{max} \leq 2+ |{\cal
  X}||{\cal Y}| \log_2 (N+1) + N \log_2|{\cal X}|$. Then $
\Pr(\sum_{i=0}^{n-1} l(\vec{\rvx}_{t-i},\vec{\rvy}_{t-i}) >
(n+\delay)NR)>0 $ only if $n>\frac{(n+\delay)NR}{l_{max}}>\frac{\delay
  NR}{l_{max}}\stackrel{\Delta}{=} \beta \delay$. So
\begin{eqnarray}
 \Pr(\vec{x}_{t}\neq \vec{x}_t((t+\delay)N ))
 &\leq & \sum_{n= \beta \delay}^{t}  \Pr(\sum_{i=0}^{n-1}
l(\vec{\rvx}_{t-i},\vec{\rvy}_{t-i})>(n+\delay)NR)
\label{eqn:univ_source_sum}
\\
& \leq_{(a)}&  \sum_{n= \beta \delay}^{t}  K_1
2^{-nN(E^{u}_{si,b}(\frac{(n+\delay)NR}{nN})-\epsilon_1)}
\nonumber\\
&\leq_{(b)} &\sum_{n=\gamma\delay}^{\infty}  K_2  2^{- n N (
E^{u}_{si,b}(R)-\epsilon_2)}
+ \sum_{n=\beta \delay}^{\gamma \delay}  K_2 2^{- \delay N (
\min_{\alpha >1}\{\frac{E^{u}_{si,b}(\alpha R)}{\alpha-1} \}-\epsilon_2)} \nonumber\\
& \leq_{(c)}&  K_3  2^{-  \gamma  \delay N (
E^{u}_{si,b}(R)-\epsilon_2)}
+ |\gamma \delay- \beta \delay|  K_3  2^{- \delay N (E_{ei}(R)-\epsilon_2)} \nonumber\\
&\leq_{(d)}& K 2^{-\delay N (E_{ei}(R)-\epsilon)}\nonumber
\end{eqnarray}
where the large $K_i's$ and arbitrarily tiny $\epsilon_i's $ are
properly chosen real numbers. $(a)$ is true because of
Lemma~\ref{lemma:source_atyp_Source_ei}.  Letting
$\gamma=\frac{E_{ei}(R)}{E^{u}_{si,b}(R)}$ in the first part of $(b)$, we
only need the fact that $E^{u}_{si,b}(R)$ is non-decreasing with $R$. In
the second part of $(b)$, let $\alpha=\frac{n+\delay}{n} $ and choose
the $\alpha$ to minimize the error exponents.  The first term of $(c)$
comes from the sum of a geometric series. The second term of $(c)$
follows from the definition of $E_{ei}(R)$ in
(\ref{eqn:focusingsource}).  $(d)$ follows from the definition of
$\gamma$ above and by absorbing the linear term into the $\epsilon$ in
the exponent. \hfill $\blacksquare$

\subsection{Converse }\label{subsec_ei_upper}

The idea is to bound the best possible error exponent with fixed
delay, without making any assumptions on the implementation of the
encoder and decoder beyond the fixed end-to-end delay constraint. In
particular, no assumption is made that the encoder works by encoding
source symbols in small groups and then uses a queue to smooth out the
rate. Instead, an encoder/decoder pair is considered that uses the
fixed-delay system to construct a fixed-block-length system. The
block-coding bounds of \cite{csiszarkorner} are thereby translated to
the fixed delay context. The arguments are analogous to the
``uncertainty-focusing bound'' derivation in \cite{OurUpperBoundPaper}
for the case of channel coding with feedback and the techniques
originate in the convolutional code literature \cite{viterbi}. 

\proof For simplicity of exposition, we ignore integer effects arising
from the finite nature of $\delay$ and $R$ etc. For every $\alpha > 0$
and delay $\delay$, consider a code running at fixed-rate till time
$\frac{\delay}{\alpha }+ \delay$. By this time, the decoder has
committed to estimates for the source symbols up to time $i =
\frac{\delay}{\alpha}$. The total number of bits generated by the
encoder during this period is $( \frac{ \delay}{\alpha }+ \delay)R$.

Now, relax the causality constraint at the encoder by giving it access
to the first $i$ source symbols all at once at the beginning of time,
rather than forcing the encoder to get the source symbols
gradually. Simultaneously, loosen the deadlines at the decoder to only
demand correct estimates for the first $i$ source symbols by the time
$ \frac{\delay}{\alpha }+ \delay$. In effect, the deadline for
decoding the {\em past} source symbols is extended to the deadline of
the $i$-th symbol itself.

Any lower-bound to the symbol error probability of the new problem is
clearly also a bound for the original problem. The difference between
block error probability and symbol error probability is at most a
factor of $\frac{1}{i}$ and is insignificant on the exponential
scale. Furthermore, the new problem is just a fixed-block-length
source coding problem requiring the encoding of $i$ source symbols
into $ ( \frac{\delay}{\alpha }+ \delay) R$ bits. The rate per symbol
is
\begin{eqnarray*}
((\frac{\delay}{\alpha }+ \delay )R)\frac{1}{i} & = &
((\frac{\delay}{\alpha }+ \delay )R) \frac{\alpha }{\delay} \\
& = & ( \alpha +1 )R.
\end{eqnarray*}

Theorem~2.15 in \cite{csiszarkorner} tells us that such a code has a
probability of error that is at least exponential in $i
E_{ei,b}((\alpha+1) R)$. Since $i = \frac{\delay}{\alpha}$, this
translates into an error exponent of at most
$\frac{E_{ei,b}((\alpha+1) R)}{\alpha}$ with parameter $\delay$.

Since this is true for all $\alpha>0$, we have the
uncertainty-focusing bound on the reliability function $E_{ei}(R)$
with fixed delay $\delay$: 
\begin{eqnarray}\label{eqn:upperbound_ei}
E_{ei}(R) \leq \inf_{\alpha>0} \frac{1}{\alpha}E_{ei,b}((\alpha
+1)R)
\end{eqnarray}
The minimizing $\alpha$ tells how much of the past
($\frac{\delay}{\alpha}$) is involved in the dominant error event.

The source uncertainty-focusing bound can be expressed parametrically
in terms of the Gallager function $E_0(\rho)$ from
(\ref{eqn:gallagerfunction}) and its slope computed in the vicinity of
the conditional entropy. This is shown in
Appendix~\ref{sec:appendix_parameterization}. \hfill $\blacksquare$

\section{No side-information at the encoder}\label{sec:source_si}

This section proves the upper bound given by
Theorem~\ref{THM_UPPER_SI} for the fixed-delay error exponent for
source coding without encoder side-information. This bound is valid
for any generic joint distribution $p_{\rvx\rvy}$. The results are
specialized to the symmetric case in
Corollary~\ref{corollary_SI_upper_symmetric}, proved in
Appendix~\ref{app:corsisymmetric}. 

\begin{figure}[htbp]
\setlength{\unitlength}{0.6mm}
\begin{picture}(100,40)(-60,0)

\put(40,30){\line(1,0){26}} \put(40,40){\line(1,0){26}}
\put(40,30){\line(0,1){10}} \put(66, 30){\line(0,1){10}}
\put(42,34){Encoder }

\put(90,30){\line(1,0){26}} \put(90,40){\line(1,0){26}}
\put(90,30){\line(0,1){10}} \put(116, 30){\line(0,1){10}}
\put(92,34){Decoder }

\put(116, 34){\vector(1,0){10}} \put(66, 34){\vector(1,0){24}}

\put(30, 5){\line(1,0){70}} \put(100, 5){\vector(0,1){25}}
\put(125,34){ $\hat{\rvx}_1,\hat{\rvx}_2,...$}

\put(30, 35){\vector(1,0){10}}

\put(0, 34){ ${\rvx}_1,{\rvx}_2,...$} \put(0,4){
${\rvy}_1,{\rvy}_2,...$}

\put(0, 18){ DMC} \put(10, 20){\oval(35 ,8) }
\put(13,32){\vector(0,-1){8}} \put(13,16){\vector(0,-1){8}}
 \end{picture}
     \caption[]{Lossless source coding with side-information only at
       the decoder.}
     \label{fig.swnewmodel}
 \end{figure}
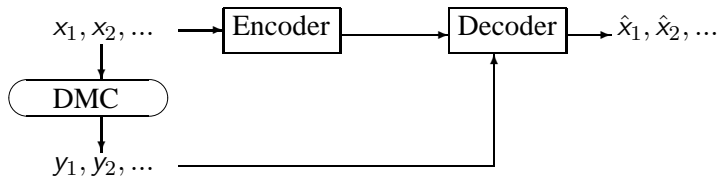

In the following analysis, it is conceptually useful to factor the
joint probability to treat the source as a random variable $\rvx$ and
consider the side-information $\rvy$ as the output of a discrete
memoryless channel (DMC) $p_{\rvy|\rvx}$ with $\rvx$ as input. This
model is shown in Figure~\ref{fig.swnewmodel}.

 The theorem is proved using a variation of the bounding technique
 used in \cite{OurUpperBoundPaper} (and originating in
 \cite{PinskerNoFeedback}) for the fixed-delay channel coding problem.
 Lemmas~\ref{lemma.Feed-forward-typeI}-\ref{lemma.ratiobound} are the
 source coding counterparts to Lemmas~4.1-4.4 in
 \cite{OurUpperBoundPaper}.  The idea of the proof is to assume a more
 powerful source decoder that has access to the previous source
 symbols (considered as feed-forward information) in addition to the
 encoded bits and the side-information. The second step is to
 construct a fixed-block-length source-coding scheme from the encoder
 and optimal feed-forward decoder. The third step is to show that if
 the side-information behaves atypically enough, then the decoding
 error probability will be large for many of the source
 symbols. The fourth step is to show that it is only future
 atypicality of the side-information that matters. This is because the
 feed-forward information allows the decoder to safely ignore all
 side-information concerning the source symbols that it already
 knows perfectly. The last step is to lower bound the probability of
 the atypical behavior and upper bound the error exponents. The proof
 spans the next several subsections.

\setlength{\unitlength}{2668sp}%

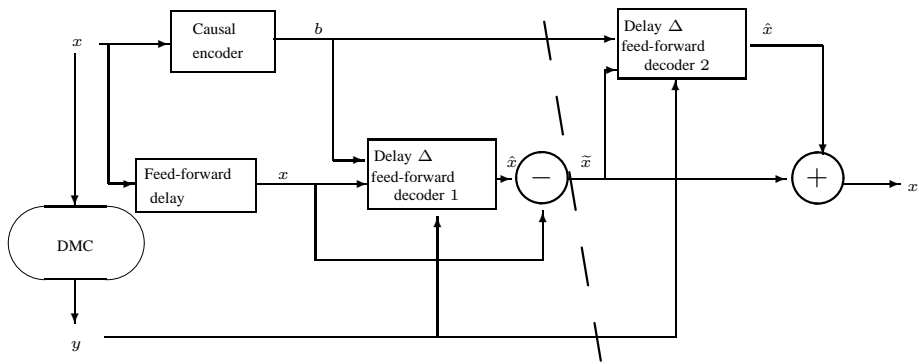
\begin{figure*}
\begin{center}
\begingroup\makeatletter\ifx\SetFigFont\undefined%
\gdef\SetFigFont#1#2#3#4#5{%
  \reset@font\fontsize{#1}{#2pt}%
  \fontfamily{#3}\fontseries{#4}\fontshape{#5}%
  \selectfont}%
\fi\endgroup%
\begin{picture}(8463,3307)(527,-3244)
\put(2251,-173){\makebox(0,0)[lb]{\smash{{\SetFigFont{6}{7.2}{\rmdefault}{\mddefault}{\updefault}Causal}}}}
\put(1801,-1523){\makebox(0,0)[lb]{\smash{{\SetFigFont{6}{7.2}{\rmdefault}{\mddefault}{\updefault}Feed-forward}}}}
\put(1801,-1748){\makebox(0,0)[lb]{\smash{{\SetFigFont{6}{7.2}{\rmdefault}{\mddefault}{\updefault}
delay}}}} \thicklines
\multiput(5513,-61)(116.72062,-700.32374){5}{\line( 1,-6){ 59.145}}
\thinlines \put(7426,-286){\line( 1, 0){562}} \put(7988,-286){\line(
1, 0){113}} \put(8101,-286){\vector( 0,-1){1012}}
\put(5738,-1523){\vector( 1, 0){2025}} \put(5738,-1523){\line( 1,
0){338}} \put(6076,-1523){\line( 0, 1){1012}}
\put(6076,-511){\vector( 1, 0){112}}
\put(988,-2198){\makebox(0,0)[lb]{\smash{{\SetFigFont{6}{7.2}{\rmdefault}{\mddefault}{\updefault}DMC}}}}
\put(1126,-286){\makebox(0,0)[lb]{\smash{{\SetFigFont{6}{7.2}{\rmdefault}{\mddefault}{\updefault}$x$}}}}
\put(3376,-173){\makebox(0,0)[lb]{\smash{{\SetFigFont{6}{7.2}{\rmdefault}{\mddefault}{\updefault}$b$}}}}
\put(3038,-1523){\makebox(0,0)[lb]{\smash{{\SetFigFont{6}{7.2}{\rmdefault}{\mddefault}{\updefault}$x$}}}}
\put(5163,-1411){\makebox(0,0)[lb]{\smash{{\SetFigFont{6}{7.2}{\rmdefault}{\mddefault}{\updefault}$\hat{x}$}}}}
\put(5851,-1411){\makebox(0,0)[lb]{\smash{{\SetFigFont{6}{7.2}{\rmdefault}{\mddefault}{\updefault}$\widetilde{x}$}}}}
\put(7538,-173){\makebox(0,0)[lb]{\smash{{\SetFigFont{6}{7.2}{\rmdefault}{\mddefault}{\updefault}$\hat{x}$}}}}
\put(8888,-1636){\makebox(0,0)[lb]{\smash{{\SetFigFont{6}{7.2}{\rmdefault}{\mddefault}{\updefault}$x$}}}}
\put(1113,-3098){\makebox(0,0)[lb]{\smash{{\SetFigFont{6}{7.2}{\rmdefault}{\mddefault}{\updefault}$y$}}}}
\put(2251,-398){\makebox(0,0)[lb]{\smash{{\SetFigFont{6}{7.2}{\rmdefault}{\mddefault}{\updefault}encoder}}}}
\put(2045,-529){\framebox(962,562){}} \put(1151,-361){\vector(
0,-1){1416}} \put(1464,-274){\line( 0,-1){1296}}
\put(1464,-1570){\vector( 1, 0){240}} \put(2839,-1574){\vector( 1,
0){1032}} \put(1376,-274){\vector( 1, 0){648}}
\put(1726,-1829){\framebox(1100,487){}} \put(1426,-2999){\line( 1,
0){3096}} \put(4522,-2999){\vector( 0, 1){1128}}
\put(5064,-1524){\vector( 1, 0){144}} \put(3001,-224){\vector( 1,
0){3168}} \put(3551,-224){\line( 0,-1){1128}}
\put(3551,-1352){\vector( 1, 0){288}} \put(3389,-1586){\line(
0,-1){696}} \put(3389,-2282){\line( 1, 0){2112}}
\put(5501,-2282){\vector( 0, 1){456}} \put(4501,-2999){\line( 1,
0){2232}} \put(6733,-2999){\vector( 0, 1){2400}}
\put(6207,-587){\framebox(1175,638){}}
\put(3882,-1799){\framebox(1175,638){}} \put(8289,-1574){\vector( 1,
0){532}} \put(1161,-2124){\oval(1252,676)} \thicklines
\put(8044,-1525){\oval(472,472)} \thinlines
\put(1151,-2461){\vector( 0,-1){408}} \thicklines
\put(5481,-1524){\oval(472,472)}
\put(3932,-1361){\makebox(0,0)[lb]{\smash{{\SetFigFont{6}{7.2}{\rmdefault}{\mddefault}{\updefault}Delay
$\Delta$}}}}
\put(6257,-149){\makebox(0,0)[lb]{\smash{{\SetFigFont{6}{7.2}{\rmdefault}{\mddefault}{\updefault}Delay
$\Delta$}}}}
\put(7945,-1586){\makebox(0,0)[lb]{\smash{{\SetFigFont{11}{12.4}{\rmdefault}{\mddefault}{\updefault}$+$}}}}
\put(5382,-1585){\makebox(0,0)[lb]{\smash{{\SetFigFont{11}{12.4}{\rmdefault}{\mddefault}{\updefault}$-$}}}}
\put(3907,-1549){\makebox(0,0)[lb]{\smash{{\SetFigFont{6}{7.2}{\rmdefault}{\mddefault}{\updefault}feed-forward}}}}
\put(4120,-1724){\makebox(0,0)[lb]{\smash{{\SetFigFont{6}{7.2}{\rmdefault}{\mddefault}{\updefault}decoder
$1$}}}}
\put(6232,-337){\makebox(0,0)[lb]{\smash{{\SetFigFont{6}{7.2}{\rmdefault}{\mddefault}{\updefault}feed-forward}}}}
\put(6445,-512){\makebox(0,0)[lb]{\smash{{\SetFigFont{6}{7.2}{\rmdefault}{\mddefault}{\updefault}decoder
$2$}}}}
\end{picture}%
 \caption[]{A cutset illustration of the Markov Chain $\rvx_1^n \ \ -\ \ (\rvxtil_1^n,\rvb_1^{\lfloor (n+\delay)R\rfloor},\rvy_1^{n+\delay})  \ \ -\ \ \rvx_1^n$.  Decoder $1$ and decoder $2$ are type I and II delay $\delay$ rate $R$ feed-forward decoders respectively.}
    \label{fig:Block_coder}
\end{center}
\end{figure*}

\subsubsection{Feed-forward decoders }
\begin{definition} A delay $\delay$ rate $R$ decoder
  $\mathcal{D}^{\delay,R}$ with feed-forward is a decoder
  $\mathcal{D}_j^{\delay,R}$ that also has access to the past source
  symbols $x_1^{j-1}$ in addition to the encoded bits
  $b_1^{\lfloor(j+\delay)R\rfloor}$ and side-information $y_1^{j+\delay}$.
\end{definition}
Using this feed-forward decoder, the estimate of $x_j$ at time
$j+\delay$ is :
\begin{eqnarray}
\xhat_{j}(j+\delay)=
\mathcal{D}_j^{\delay,R}(b_1^{\lfloor(j+\delay)R\rfloor},
y_1^{j+\delay},x_1^{j-1} )
\end{eqnarray}

\begin{lemma}{}\label{lemma.Feed-forward-typeI} For any rate $R$
  encoder $\mathcal{E}$, the optimal delay $\delay$ rate $R$ 
decoder $\mathcal{D}^{\delay,R}$ with feed-forward only needs to
depend on $b_1^{\lfloor(j+\delay)R\rfloor},
y_j^{j+\delay},x_1^{j-1}$
\end{lemma}
\proof The source and side-information pair $(\rvx_i, \rvy_i)$ is an
iid random process and the encoded bits $b_1^{\lfloor(j+\delay)R\rfloor}$
are causal functions of $x_1^{j+ \delay}$. It is easy to see that the
Markov chain 
$y_1^{j-1} \ \ -\ \
(x_1^{j-1},b_1^{\lfloor(j+\delay)R\rfloor},y_j^{j+\delay} ) \ \ -\ 
\ x_{j}^{j+ \delay}$ holds since
\begin{eqnarray*}
& & \Pr(x_{j}^{j+
  \delay},y_1^{j-1},x_1^{j-1},b_1^{\lfloor(j+\delay)R\rfloor},y_j^{j+\delay})
\\
& = & 
\Pr(x_1^{j-1},b_1^{\lfloor(j+\delay)R\rfloor},y_j^{j+\delay})
\Pr(x_{j}^{j+
  \delay}|x_1^{j-1}, b_1^{\lfloor(j+\delay)R\rfloor},y_j^{j+\delay})
\Pr(y_1^{j-1}|x_1^{j-1},
b_1^{\lfloor(j+\delay)R\rfloor},y_j^{j+\delay}, x_{j}^{j+ \delay}) \\
& = & 
\Pr(x_1^{j-1}, b_1^{\lfloor(j+\delay)R\rfloor}, y_j^{j+\delay})
\Pr(x_{j}^{j+ \delay} |x_1^{j-1}, b_1^{\lfloor(j+\delay)R\rfloor}, y_j^{j+\delay})
\Pr(y_1^{j-1}|x_1^{j-1})
\end{eqnarray*}
Thus, conditioned on the past source symbols, the
past side-information is completely irrelevant for optimal MAP 
estimation of $x_{j}$.\hfill$\square$

Write the error sequence of the feed-forward decoder as $\xtil_i=
\svx_i - \xhat_i$ by identifying the finite source alphabet with the
appropriate finite group. Then we have the following property for the
feed-forward decoders.

\begin{lemma}{}\label{lemma.Feed-forward-typeII} Given a rate $R$
encoder $\mathcal{E}$, the optimal delay $\delay$ rate $R$ decoder
$\mathcal{D}^{\delay,R}$ with feed-forward for symbol $j$ only needs
to depend on $b_1^{\lfloor(j+\delay)R\rfloor},
y_1^{j+\delay},\xtil_1^{j-1}$
\end{lemma}

\proof Proceed by induction. It holds for $j=1$ since there are no
prior source symbols. Suppose that it holds for all $j<k$ and
consider $j=k$. By the induction hypothesis, the action of all the
prior decoders $j$ can be simulated using
$(b_1^{\lfloor(j+\delay)R\rfloor}, y_1^{j+\delay},\xtil_1^{j-1})$
giving $\xhat_1^{k-1}$. This in turn allows the recovery of
$x_1^{k-1}$ since we also know $\xtil_1^{k-1}$. Thus the optimal
feed-forward decoder can be expressed in this form. \hfill$\square$

We call the feed-forward decoders in Lemmas
\ref{lemma.Feed-forward-typeI} and \ref{lemma.Feed-forward-typeII}
type I and II delay $\delay$ rate $R$ feed-forward decoders
respectively.  Lemmas \ref{lemma.Feed-forward-typeI} and
\ref{lemma.Feed-forward-typeII} tell us that feed-forward decoders can
be thought in three ways: having access to all encoded bits, all side
information and all past source symbols,
$(b_1^{\lfloor(j+\delay)R\rfloor}, y_1^{j+\delay},x_1^{j-1} )$, having
access to all encoded bits, a recent window of side information and
all past source symbols, $(b_1^{\lfloor(j+\delay)R\rfloor},
y_j^{j+\delay},x_1^{j-1})$, or having access to all encoded bits, all
side information and all past decoding errors,
$(b_1^{\lfloor(j+\delay)R\rfloor}, y_1^{j+\delay},\xtil_1^{j-1} )$.

\subsubsection{Constructing a block code } To encode a block of $n$
source symbols, just run the rate $R$ encoder $\mathcal{E}$ and
terminate with the encoder run using some $\delay$ random source
symbols drawn according to the distribution of $p_\rvx$. To decode the
block, just use the delay $\delay$ rate $R$ decoder
$\mathcal{D}^{\delay,R}$ with feed-forward, and then further use the
fedforward error signals to correct any mistakes that might have
occurred. As a block coding system, this hypothetical system never
makes an error from end to end. As shown in
Figure~\ref{fig:Block_coder}, the data processing inequality implies:

\begin{lemma} \label{lemma.entropy-bound} If $n$ is the fixed block-length,
  and the block rate is $R (1+\frac{\delay}{n})$, then
\begin{eqnarray}\label{eqn.block_entropy}
H(\rvxtil_1^n) \geq  - (n+\delay)R +n H(\rvx|\rvy)
\end{eqnarray}
\end{lemma}

\proof: See Appendix~\ref{app:lementropybound}.

\subsubsection{ Lower bound the symbol-wise error probability }

Now suppose this block-code were to be run with the distribution
$q_{\rvx\rvy}$, s.t.~$H(q_{\rvx|\rvy})> (1+\frac{\delay}{n})R$, from
time $1$ to $n$, and were to be run with the distribution
$p_{\rvx\rvy}$ from time $n+1$ to $n+\delay$. Write the hybrid
distribution as $Q_{\rvx\rvy}$. Then the block coding scheme
constructed in the previous section would with probability very close
to $1$ make a block error. Moreover, many individual symbols would
also be in error often:

\begin{lemma} \label{lemma.errorprobabilitybound}
If the source and side-information is coming from $q_{\rvx\rvy}$, then
there exists a $\delta > 0$ so that for $n$ large enough, there exists
a number $n_e$ and a sequence of symbol positions
$j_1 < j_2 < \ldots < j_{n_e}$ satisfying:
\begin{itemize}
 \item $n_e \geq \frac{H(q_{\rvx|\rvy})
     -\frac{n+\Delta}{n}R}{2\log_2|\mathcal{X}|- (H(q_{\rvx|\rvy})
     -\frac{n+\Delta}{n}R)}n$ 
 \item The probability of symbol errors made by the feed-forward
       decoder on symbol $\rvx_{j_i}$ is at least $\delta$ when the
       joint source symbols are drawn according to $q_{\rvx\rvy}$.

 \item $\delta$ satisfies
$h_\delta+\delta\log_2(|\mathcal{X}|-1) =\frac{1}{2}(
H(q_{\rvx|\rvy}) -\frac{n+\Delta}{n}R)$ where
$h_\delta=-\delta\log_2\delta-(1-\delta)\log_2(1-\delta)$. 
\end{itemize}
\end{lemma}
 \proof See Appendix~\ref{app:lemerrorprobbound}.
\vspace{0.1in}

Pick $j^*=j_{\frac{n_e}{2}}$ to pick a symbol position in the middle
of the block that is subject to
errors. Lemma~\ref{lemma.errorprobabilitybound} reveals that
$\min\{j^* , n- j^*\}\geq \frac{1}{2} \frac{ (H(q_{\rvx|\rvy})
  -\frac{n+\Delta}{n}R) }{ 2\log_2|\mathcal{X}|-(H(q_{\rvx|\rvy})
  -\frac{n+\Delta}{n}R)}n$, so if we fix $\frac{\delay}{n}$ and let
$n$ go to infinity, then $\min\{j^* , n- j^*\} $ goes to infinity as
well.

At this point, Lemma~\ref{lemma.Feed-forward-typeI} implies that the
decoder can ignore the side-information from the past. Define the
``bad sequence'' set $E_{j^*}$ as the set of source and
side-information sequence pairs so the type I delay $\delay$ rate $R$
decoder makes a symbol error at $j^*$. To simplify notation, let
$\vec{x}=x_1^{j^*+\delay}$, $\xBar=x_1^{j^*-1}$,
$\svxBBar=x_{j^*}^{j^*+\delay}$, $\svyBBar=y_{j^*}^{j^*+\delay}$
denote the entire source vector, the source prefix, and the suffixes
for the source and side-information respectively.  
Define $E_{j^*}=\{(\vec{x}, \svyBBar)| x_{j^*}\neq
\mathcal{D}_{j^*}^{\delay,R}(\mathcal{E}(\vec{x}),
y_j^{j^*+\delay},\xBar) \} $. 

Since the ``bad sequence'' set $E_{j^*}$ only has future
side-information in it, the probability of this set depends only on
the marginals for $\rvx$ in the past and the joint distribution in the
present and future. Consider a hybrid distribution where the joint
source behaves according to $Q_{\rvx\rvy}$ from time $j^*$ to
$j^*+\delay$ \textit{and} the $\rvx$ source one behaves like it came
from a distribution $q_\rvx$ from time $1$ to $j^*-1$. By
Lemma~\ref{lemma.errorprobabilitybound}, $Q_{\rvx\rvy}(E_{j^*}) \geq
\delta$. 

Define $J =\min \{n, j^*+\delay\}$ to deal with possible
edge-effects\footnote{These edge effects, although annoying, cannot be
  ignored since guaranteeing that $\delta$ is small relative to $n$
  would come at the cost of less tight bounds in asymmetric cases. In
  this way, the situation is different from the argument given in
  \cite{OurUpperBoundPaper} for channel coding without feedback.}
near the end of the block, and let $\bar{\svxBBar}=x_{j^*}^{J}$,
$\bar{\svyBBar}=y_{j^*}^{J}$. The empirical distribution of
$(\svxBBar,\svyBBar)$ is written using shorthand
$r_{\svxBBar,\svyBBar}(x,y)=\frac{n_{x,y}(\svxBBar,\svyBBar)}{\delay+1}$
and similarly the empirical distribution of $\xBar$ as $r_{\xBar
}(x)=\frac{n_{x}(\xBar )}{j^*-1}$.

Now, the strongly typical set can be defined
\begin{eqnarray}
A^{\epsilon}_{J}(q_{\rvx\rvy})= \{&&(\vec{x},\svyBBar)\in
\mathcal{X}^{j^*+\delay}\times \mathcal{Y}^{\delay+1}| \forall x,
r_{\xBar}(x)\in(q_{\rvx}(x)-\epsilon,
q_{\rvx}(x)+\epsilon)\nonumber\\
&& \forall (x,y)\in \mathcal{X}\times\mathcal{Y}, \ \
q_{\rvx\rvy}(x,y)>0 :
r_{\bar{\svxBBar},\bar{\svyBBar}}(x,y)\in(q_{\rvx\rvy}(x,y)-\epsilon,
q_{\rvx\rvy}(x,y)+\epsilon),\nonumber\\
&&\forall (x,y)\in \mathcal{X}\times\mathcal{Y},\ \
q_{\rvx\rvy}(x,y)=0 : r_{\bar{\svxBBar},\bar{\svyBBar}}(x,y)=0
 \ \ \ \}. \label{eqn:typicaldef}
\end{eqnarray}
The conditions require that the prefix be $q_{\rvx}$-typical and the
suffix till $J$ be $q_{\rvx\rvy}$-typical. What happens after $J$ is
not important.

This typical set is used to get a sequence of lemmas asserting that
errors are common even when we restrict to the typical behavior of the
$q$ distribution, that the probability of $q$-typical joint
realizations is least exponentially small under the true distribution,
and that this means that the errors themselves must occur at least
with exponentially small probability.

\begin{lemma} \label{lemma.typicalset} $Q_{\rvx\rvy}(E_{j^*}\cap
  A^{\epsilon}_{J}(q_{\rvx\rvy})) \geq \frac{\delta}{2}$ for large $n$
  and $\delay$. 
\end{lemma}

\proof See Appendix~\ref{app:lemtypicalset}.
\vspace{0.1in}

\begin{lemma} \label{lemma.ratiobound}  For all  $\epsilon< \min_{x,y: p_{\rvx\rvy}(x,y)>0}\{ p_{\rvx\rvy}(x,y)\}$, $\forall (\vec{x},
\svyBBar)\in A^{\epsilon}_{J}(q_{\rvx\rvy})$,
$$\frac{p_{\rvx\rvy}(\vec{x}, \svyBBar)}{Q_{\rvx\rvy}(\vec{x},
\svyBBar)}\geq  2^{-(J-j^*+1)  D(q_{\rvx\rvy}\|p_{\rvx\rvy})
-(j^*-1)  D(q_{\rvx}\|p_{\rvx}) -JG \epsilon}$$

where
 $G=\max\{|\mathcal{X}||\mathcal{Y}|+\sum_{x,y: p_{\rvx\rvy}(x,y)>0 }
\log_2(\frac{q_{\rvx\rvy} (x,y)}{p_{\rvx\rvy}(x,y)}+1),|\mathcal{X}|
+\sum_{x} \log_2(\frac{q_{\rvx } (x )}{p_{\rvx }(x )}+1)\} $
\end{lemma}

\proof  See Appendix~\ref{app:lemratiobound}.
\vspace{0.1in}

\begin{lemma}\label{lemma.proofoftheorem} For all $ \epsilon<
\min_{x,y}\{  p_{\rvx\rvy}(x,y)\}$, and large $\delay$, $n$:
$$p_{\rvx\rvy}(E_{j^*}) \geq \frac{\delta}{2}  2^{-(J-j^*+1)  D(q_{\rvx\rvy}\|p_{\rvx\rvy}) -(j^*-1)
D(q_{\rvx}\|p_{\rvx}) -JG \epsilon}$$
\end{lemma}
\proof See Appendix~\ref{app:lemproofoftheorem}.

\subsubsection{Final details in proving Theorem~\ref{THM_UPPER_SI}}

Notice that as long as $H(q_{\rvx|\rvy})>\frac{n+\delay}{n}R$, we know
$\delta>0$ by letting $\epsilon$ go to $0$, and having $\delay$ and
$n$ (and thus also $J$) go to infinity proportionally. So $\Pr [
\rvxhat_{j^*} (j^*+\delay) \neq \rvx_{j^*} ]=
p_{\rvx\rvy}(E_{j^*})\geq K 2^{-(J-j^*+1)
  D(q_{\rvx\rvy}\|p_{\rvx\rvy}) -(j^*-1) D(q_{\rvx}\|p_{\rvx}) }$.

Notice that $D(q_{\rvx\rvy}\|p_{\rvx\rvy}) \geq D(q_{\rvx}\|p_{\rvx
})$. Since  $J=\min\{n,j^*+\delay\}$, for all possible $j^*\in
[1,n]$ we have for all $n \geq \delay$:
\begin{eqnarray}
 (J-j^*+1) D(q_{\rvx\rvy}\|p_{\rvx\rvy}) +(j^*-1)
D(q_{\rvx}\|p_{\rvx})
&\leq& (\delay+1)D(q_{\rvx\rvy}\|p_{\rvx\rvy})+(n-\delay-1)D(q_{\rvx}\|p_{\rvx})\nonumber\\
& \approx &
\delay(D(q_{\rvx\rvy}\|p_{\rvx\rvy})+\frac{n-\delay}{\delay}D(q_{\rvx}\|p_{\rvx})).\nonumber
\end{eqnarray}

Meanwhile, for $n < \delay$:
\begin{eqnarray}
(J-j^*+1) D(q_{\rvx\rvy}\|p_{\rvx\rvy}) +(j^*-1)
D(q_{\rvx}\|p_{\rvx})
 \leq  n D(q_{\rvx\rvy}\|p_{\rvx\rvy})
 = \delay(\frac{n}{\delay}D(q_{\rvx\rvy}\|p_{\rvx\rvy}) ).\nonumber
\end{eqnarray}

Write $\alpha =\frac{\delay}{n}$. The upper bound on the error
exponent is the minimum of the above error exponents over all
$\alpha>0$.

\begin{eqnarray}
 E_{si}^{u}(R)= \min && \{\inf_{q_{\rvx\rvy}, \alpha\geq 1:
H(q_{\rvx|\rvy})>(1+\alpha)R
}\{\frac{1}{\alpha}D(q_{\rvx\rvy}\|p_{\rvx\rvy}) \}, \nonumber\\
&&\inf_{q_{\rvx\rvy}, 1\geq \alpha\geq 0:
H(q_{\rvx|\rvy})>(1+\alpha)R
}\{\frac{1-\alpha}{\alpha}D(q_{\rvx}\|p_{\rvx})
+D(q_{\rvx\rvy}\|p_{\rvx\rvy}) \} \}\nonumber
\end{eqnarray}
This is the desired result. \hfill $\blacksquare$

The specialization of this result to uniform sources $\rvx$ and side
information $\rvy=\rvx\oplus\rvs$ is straightforward and is covered in
Appendix~\ref{app:corsisymmetric}. The key is to understand that when
the joint source is symmetric, the marginal for $q$ always agrees with
the marginal for the original $p$. 

\section{Conclusions} \label{sec:conclusion}

This paper has shown that fixed-block-length and fixed-delay lossless
source-coding behave very differently when decoder side-information is
either present or absent at the encoder. While fixed-block-length
systems do not usually gain substantially in reliability with encoder
access to the side-information, fixed-delay systems can achieve very
substantial gains in reliability. This means that if an application
has a target for both end-to-end latency and probability of symbol
error, then depriving the encoder of access to the side-information
will come at the cost of higher required data rates.

The proof of achievability makes clear the connection to ideas of
``effective bandwidth'' and buffer-provisioning in the networking
context (see e.g.\cite{changthomaseffective}). The results here and 
in \cite{OurUpperBoundPaper} can be considered a way to extend the
spirit of those concepts to problems like source-coding without access
to side-information and communication without
feedback. Thinking about buffer-overflow is too narrow a perspective
to generalize the idea of ``how much extra rate is required beyond the
minimum'' but end-to-end delay provides a framework to understand this
and thereby compare different approaches.

Thus, it is useful to view this paper as a companion to its sister paper
\cite{OurUpperBoundPaper} (treating channel-coding with and without
feedback) in the fixed-delay context. Comparing both sets of results
shows how feedback in channel coding is very much like encoder access
to decoder side-information in lossless source coding. The main
difference is that source coding performance is generally better at
high rates while channel coding is better at low rates. The subtle
aspect of the analogy is that lossless source-coding with encoder
side-information behaves like channel-coding with feedback {\em for
  channels with strictly positive zero-error capacity}.

\begin{itemize}
\item For generic symmetric channels with $C_{0,f} > 0$, the
  fixed-block-length reliability function is known perfectly with
  feedback and jumps abruptly to $\infty$ at $C_{0,f}$ and approaches
  zero quadratically at $C$.

  For generic sources, the fixed-block-length reliability function is
  known perfectly with encoder side-information and jumps abruptly
  to $\infty$ at $\log_2 |{\cal X}|$ and approaches zero quadratically
  at $H_{\rvx|\rvy}$.

\item For generic symmetric channels with $C_{0,f} > 0$, the
  fixed-delay reliability with feedback tends smoothly to $\infty$ at
  $C_{0,f}$ and approaches zero linearly at $C$.

  For general sources with encoder access to side-information, the
  fixed-delay reliability function tends smoothly to $\infty$ at
  $\log_2 |{\cal X}|$ and approaches zero linearly at $H_{\rvx|\rvy}$.

\item For generic symmetric channels with $C_{0,f} > 0$, an
  asymptotically optimal fixed-delay code with feedback can be
  constructed using a queue fed at fixed rate followed by a
  fixed-to-variable channel code. 

  For generic sources with encoder access to side-information, an
  asymptotically optimal fixed-delay code can be constructed using a
  fixed-to-variable source code followed by a queue drained at fixed
  rate.
\end{itemize}

In both cases, the non-ignorant encoders can help deliver
substantially lower end-to-end delays. In addition, in both cases
there is a gap between the achievable regions and converses for fixed
delay reliability for asymmetric cases when considering ignorant
encoders. In addition to closing this gap, many natural problems
remain to be explored: joint source-channel coding
\cite{Allerton06WithCheng}, lossy coding \cite{ISIT07WithCheng}, as
well as extending the upper-bound techniques here to truly
multi-terminal settings with distributed encoders.

\appendices

\section{Proof of Lemma~\ref{lemma:source_atyp_Source_ei}} \label{app:achievabilityqueue} 
In the large deviation theory literature, limit superior and limit inferior
are widely used while calculating the asymptotic properties of the
rate functions \cite{DemboZeitouni}. However it is sometimes more
convenient to use the following equivalent $\epsilon-K$ conditions
since 
\begin{eqnarray}
a\leq \liminf_{n\rightarrow\infty} \frac{1}{n}\log_2 P_n\leq
\limsup_{n\rightarrow\infty} \frac{1}{n}\log_2 P_n \leq b\nonumber
\end{eqnarray}
iff for all $\epsilon>0$, there exists $K< \infty$, such
that for all $n$: $ K 2^{n(a-\epsilon)} \leq P_n \leq K
2^{n(b+\epsilon)} $. The equivalence is obvious from the definitions
of limit superior and limit inferior \cite{DemboZeitouni}. 

\proof By Cram\'{e}r's theorem\cite{DemboZeitouni}, for all
$\epsilon_1>0$ there exists $K_1$, such that: 
\begin{eqnarray}
 \Pr(\sum_{i=1}^n l(\vec{\rvx}_i, \vec{\rvy}_i) > nN r )=
\Pr(\frac{1}{n}\sum_{i=1}^n l(\vec{\rvx}_i, \vec{\rvy}_i) > N r )
 \leq  K_1 2^{-n (\inf_{z > N
r}I(z)-\epsilon_1)}\label{eqn:ratefunctionSingleSource_ei}
\end{eqnarray}
where the rate function $I(z)$ is \cite{DemboZeitouni}:
\begin{eqnarray}
I(z) &=&\sup_{\rho \in \mathcal{R}}\{ \rho z-\log_2(
\sum_{(\vec{\svx}, \vec{\svy})\in \mathcal{X}^N\times \mathcal{Y}^N}
p_{\rvx\rvy}(\vec{x},\vec{y})2^{\rho
l(\vec{x},\vec{y})})\}\label{eqn:definition_of_I(z)_ei}
\end{eqnarray}

$$\mbox{Write} \ \ \ I(z,\rho)= \rho z-\log_2( \sum_{(\vec{\svx}, \vec{\svy})\in
\mathcal{X}^N\times \mathcal{Y}^N}
p_{\rvx\rvy}(\vec{x},\vec{y})2^{\rho l(\vec{x},\vec{y})})$$
Notice that the H\"{o}lder inequality implies that
for all $\rho_1,\rho_2$ and for all $\theta\in (0,1)$:
\begin{eqnarray}
( \sum_i p_i 2^{\rho_1 l_i})^\theta   ( \sum_i p_i 2^{\rho_2
l_i})^{(1-\theta)}
&\geq& \sum_i (p_i^\theta 2^{\theta\rho_1 l_i})(p_i^{(1-\theta)} 2^{(1-\theta)\rho_2 l_i}) \nonumber\\
&=&  \sum_i p_i 2^{(\theta\rho_1+(1-\theta)\rho_2) l_i}\nonumber
\end{eqnarray}
This shows that $\log_2( \sum_{(\vec{\svx}, \vec{\svy})\in
  \mathcal{X}^N\times \mathcal{Y}^N}
p_{\rvx\rvy}(\vec{x},\vec{y})2^{\rho l(\vec{x},\vec{y})})$ is a convex
$\cup$ function of $\rho$ and thus $I(z,\rho)$ is a concave $\cap$
function of $\rho$ for fixed $z$. Clearly $I(z,0)=0$. Consider $z > Nr
>NH(\rvx|\rvy)$. For large $N$,
\begin{eqnarray}
\frac{\partial I(z,\rho)}{\partial\rho}|_{\rho=0} =
z-\sum_{(\vec{\svx}, \vec{\svy})\in \mathcal{X}^N\times
\mathcal{Y}^N} p_{\rvx\rvy}(\vec{x},\vec{y}) l(\vec{x}, \vec{y})\geq
0\nonumber
\end{eqnarray}
since the average codeword length is essentially $NH(\rvx|\rvy)$. Thus
$I(z,\rho)<0$ as long as $z> Nr$ and $\forall \rho<0$.  This means
that the $\rho$ to maximize $I(z,\rho)$ is positive. So from now on,
it is safe to assume $\rho\geq 0$.  This implies that $I(z)$ is
monotonically increasing with $z$ and it is obvious that $I(z)$ is
continuous. Thus

\begin{eqnarray}
\inf_{z> Nr}I(z) = I(Nr).\label{eqn:ei_temp1}
\end{eqnarray}
Using the upper bound on  $l(\vec{x}, \vec{y})$ in
(\ref{eqn:lower_upperboundonl(s)_ei}):
\begin{eqnarray}
\log_2( \sum_{(\vec{\svx}, \vec{\svy})\in \mathcal{X}^N\times
\mathcal{Y}^N} p_{\rvx\rvy}(\vec{x},\vec{y})2^{\rho
l(\vec{x},\vec{y})})
&\leq& \log_2( \sum_{q_{\rvx\rvy}\in \mathcal{T}^N }
2^{-ND(q_{\rvx\rvy}\|p_{\rvx\rvy})}2^{\rho(\epsilon_N + NH(q_{\rvx|\rvy}))  })\nonumber\\
& \leq & 2^{N \epsilon_N} 2^{-N\min_q
\{D(q_{\rvx\rvy}\|p_{\rvx\rvy})- \rho
H(q_{\rvx|\rvy})  -\rho \epsilon_N \}   })\nonumber\\
 &=&N\big(- \min_q \{D(q_{\rvx\rvy}\|p_{\rvx\rvy})- \rho H(q_{\rvx|\rvy})-\rho\epsilon_N\}+\epsilon_N\big)\nonumber
\end{eqnarray}
where $ 0<\epsilon_N\leq\frac{2 + |{\cal X}||{\cal Y}| \log_2 (N+1)}{N}$
goes to $0$ as $N$ goes to infinity and $\mathcal{T}^N$ is the set of
all joint types of $\mathcal{X}^N\times \mathcal{Y}^N$.

Substitute the above inequalities into $I(Nr)$ defined in
(\ref{eqn:definition_of_I(z)_ei}):
\begin{eqnarray}
I(Nr)\geq N \big( \sup_{\rho\geq 0}\{\min_q \rho
(r-H(q_{\rvx|\rvy})-\epsilon_N) +D
(q_{\rvx\rvy}\|p_{\rvx\rvy})\}-\epsilon_N
\big)\label{eqn:INrexpression_ei}
\end{eqnarray}

The next task is to show that $I(Nr) \geq N(E_{ei,b}(r)+\epsilon)$
where $\epsilon$ goes to $0$ as $N$ goes to infinity. This can be
proved by the tedious but direct Lagrange multiplier method used in
\cite{StreamingSlepianWolf}. Instead, the proof here is based on the
existence of a saddle point. Define
$$f(q,\rho)=\rho
(r-H(q_{\rvx|\rvy})-\epsilon_N) +D (q_{\rvx\rvy}\|p_{\rvx\rvy}).$$

Clearly for fixed $q$, $f(q,\rho)$ is a linear function of $\rho$, and
thus concave. In addition, for fixed $\rho \geq 0$, $f(q,\rho)$ is a
convex $\cup$ function of $q$, because both $-H(q_{\rvx|\rvy})$ and $D
(q_{\rvx\rvy}\|p_{\rvx\rvy})$ are convex $\cup$ on
$q_{\rvx\rvy}$. Define $g(u)\triangleq \min_q \sup_{\rho\geq
  0}(f(q,\rho)+\rho u)$. It is enough to show that $g(u)$ is finite in
the neighborhood of $u=0$ to establish the existence of the saddle
point \cite{BoydConvex}.
\begin{eqnarray}
g(u)&=_{(a)}&\min_q \sup_{\rho\geq0} f(q,\rho)+\rho u \nonumber\\
&=_{(b)}& \min_q \sup_{\rho\geq0} \rho
(r-H(q_{\rvx|\rvy})-\epsilon_N+u)
+D (q_{\rvx\rvy}\|p_{\rvx\rvy})  \nonumber\\
&\leq_{(c)}& \min_{q:H(q_{\rvx|\rvy})\geq r-\epsilon_N+u}
\sup_{\rho\geq0} \rho (r-H(q_{\rvx|\rvy})-\epsilon_N+u)
+D (q_{\rvx\rvy}\|p_{\rvx\rvy})  \nonumber\\
&\leq_{(d)}& \min_{q:H(q_{\rvx|\rvy})\geq r-\epsilon_N+u} D
(q_{\rvx\rvy}\|p_{\rvx\rvy})\nonumber\\
&<_{(e)}& \infty
\end{eqnarray}
$(a)$, $(b)$ are from the definitions. $(c)$ is true because
$H(p_{\rvx|\rvy}) < r < \log_2|\mathcal{X}|$ and thus for very small
$\epsilon_N$ and $u$, $H(p_{\rvx|\rvy}) < r-\epsilon_N+u <
\log_2|\mathcal{X}|$. Consequently, there exists a distribution $q$ so
that $H(q_{\rvx|\rvy}) \geq r-\epsilon_N+u$. $(d)$ holds because
$H(q_{\rvx|\rvy}) \geq r-\epsilon_N+u$ and $\rho\geq 0$. $(e)$ is
true because we assumed without loss of generality that the marginal 
$p_\rvx(x)>0$ for all $x\in \mathcal{X}$ together with the fact that
$r-\epsilon_N+u < \log_2|\mathcal{X}|$.  The finiteness implies the
existence of the saddle point of $f(q,\rho)$.
\begin{eqnarray}
\sup_{\rho\geq 0}\{\min_q f(q,\rho)\}= \min_q\{ \sup_{\rho\geq
0}f(q,\rho)\}\label{eqn:saddlepoint}
\end{eqnarray}

Note that if $H(q_{\rvx|\rvy})< r+\epsilon_N$, then $\rho$ can be
chosen to be arbitrarily large to make $\rho
(r-H(q_{\rvx|\rvy})-\epsilon_N) +D (q_{\rvx\rvy}\|p_{\rvx\rvy})$
arbitrarily large. Thus the $q$ to minimize $\sup_\rho \rho
(r-H(q_{\rvx|\rvy})-\epsilon_N) +D (q_{\rvx\rvy}\|p_{\rvx\rvy})$
satisfies $r-H(q_{\rvx|\rvy})-\epsilon_N\geq 0$. Thus
\begin{eqnarray}
\min_q\{\sup_{\rho\geq 0} \rho (r-H(q_{\rvx|\rvy})-\epsilon_N) +D
(q_{\rvx\rvy}\|p_{\rvx\rvy})\}&=_{(a)}&\min_{q: H(q_{\rvx|\rvy})\geq
r-\epsilon_N}\sup_{\rho\geq 0} \{\rho
(r-H(q_{\rvx|\rvy})-\epsilon_N)+ D
(q_{\rvx\rvy}\|p_{\rvx\rvy})\}\nonumber\\
&=_{(b)}&\min_{q: H(q_{\rvx|\rvy})\geq r-\epsilon_N}\{ D
(q_{\rvx\rvy}\|p_{\rvx\rvy})\}\nonumber\\
&=_{(c)}& E_{ei, b}(r-\epsilon_N).\label{eqn:minmax_expression}
\end{eqnarray}

$(a)$ follows from the argument above. $(b)$ is true because
$r-H(q_{\rvx|\rvy})-\epsilon_N\leq 0$ and $\rho\geq 0$ and thus
$\rho=0$ maximizes $\rho(r-H(q_{\rvx|\rvy})-\epsilon_N)$. $(c)$ is
true by definition. Combining (\ref{eqn:INrexpression_ei})
(\ref{eqn:saddlepoint}) and (\ref{eqn:minmax_expression}),  letting
$N$ be sufficiently big implies that $\epsilon_N$ is sufficiently
small. Noticing that $E_{ei}(r)$ is continuous on $r$, we get the the
desired bound in (\ref{eqn:source_atypicality_ei}). \hfill$\square$
\vspace{0.1 in}

\section{Parametrization of $E_{ei}(R)$}\label{sec:appendix_parameterization}

We need the definition of tilted distributions for a joint
distribution $p_{\rvx\rvy}$ from \cite{StreamingSlepianWolf}. 

\begin{definition}   {$\rvx-\rvy$ tilted distribution of $p_{\rvx\rvy}$}: $\bar{p}^\rho_{\rvx\rvy}$, for all $\rho \in
[-1,+\infty)$
\begin{eqnarray}
 \bar{p}^\rho_{\rvx\rvy}(x,y) &=&\frac{[\sum_s p_{\rvx\rvy}(s,y)^{\frac{1}{1+\rho}}]^{1+\rho}}{\sum_t[\sum_s p_{\rvx\rvy}(s,t)^{\frac{1}{1+\rho}}]^{1+\rho}}\times\frac{p_{\rvx\rvy}(x,y)^{\frac{1}{1+\rho}}}{\sum_s p_{\rvx\rvy}(s,y)^{\frac{1}{1+\rho}}} \nonumber\\
\end{eqnarray}
\end{definition}
Write the conditional entropy of $\rvx$ given $\rvy$ for this tilted
distribution as $H(\bar{p}^\rho_{\rvx|\rvy})$. An important fact as
shown in Lemma~17 of \cite{StreamingSlepianWolf} is that
$\frac{\partial E_0(\rho)}{\partial \rho} =
H(\bar{p}^\rho_{\rvx|\rvy})$, also
$H(\bar{p}^\rho_{\rvx|\rvy})|_{\rho=0}=H(p_{\rvx|\rvy})$,
$H(\bar{p}^\rho_{\rvx|\rvy})|_{\rho=+\infty}=\log_2(M(p_{\rvx\rvy}))$. 
where $M(p_{\rvx\rvy})=|\max_{y\in\mathcal{Y}}\{x\in \mathcal{X}:
p_{\rvx\rvy}(x,y)>0\}|$

We first show that $\frac{E_0(\rho)}{\rho}$ is in general
monotonically increasing for $\rho\in [0,\infty)$:

\begin{eqnarray}
\frac{\partial \frac{E_0(\rho)}{\rho}}{\partial\rho}&=_{(a)}& \frac{\rho\frac{\partial E_0(\rho)}{\partial \rho}-E_0(\rho)}{\rho^2}\nonumber\\
&=_{(b)}& \frac{\rho H(\bar{p}^\rho_{\rvx|\rvy})-E_0(\rho)}{\rho^2}\nonumber\\
&=_{(c)}&
\frac{D(\bar{p}^\rho_{\rvx\rvy}\|p_{\rvx\rvy})}{\rho^2}\nonumber\\
&>_{(d)}& 0\nonumber
\end{eqnarray}
(a) is obvious and (b) is from Lemma~17 in \cite{StreamingSlepianWolf}. (c) is
from Lemma~15 in \cite{StreamingSlepianWolf}. (d) is true unless the
source $\rvx$ is conditionally uniform given side information
$\rvy$. For the trivial case conditionally uniform case where
$p_{\rvx|\rvy}(x|y)=\frac{1}{M(p_{\rvx\rvy})}$ on those letters $x$
for which it is nonzero, both the fixed-block-length error
exponent $E_{si,b}^u(R)$ and the delay 
error exponent $E_{ei}(R) $ are either $0$ when
$R<\log_2(M(p_{\rvx\rvy}))$ or $\infty$ when
$R>\log_2(M(p_{\rvx\rvy}))$. 

With the above observations, we know that for all $R\in
[H(p_{\rvx|\rvy}),\log_2 M(p_{\rvx\rvy}))$, there exists a unique
$\rho^*\geq 0$,s.t. $R = \frac{E_0(\rho^*)}{\rho^*}$ or equivalently
$\rho^* R = E_0(\rho^*)$. In order to show
(\ref{eqn:parameterization_of_Eei}), it remains to show that
$E_{ei}(R) = E_0(\rho^*)$.

 From the definition of $E_{ei}(R)$ in (\ref{eqn:focusingsource})
and the definition of $E_{si,b}^u(R)$ in
(\ref{eqn:blockupperboundrho}), we have:

\begin{eqnarray}
 E_{ei}(R)&=&\inf_{\alpha>0}  \frac{1}{\alpha}E_{si,b}^u((\alpha+1)R)\nonumber\\
 &=&\inf_{\alpha>0} \big\{\sup_{\rho\geq 0} \frac{\rho(\alpha+1)
 R-E_0(\rho)}{\alpha}\big\}\nonumber\\
 &\geq&  \sup_{\rho\geq 0}\inf_{\alpha>0} \rho R + \frac{\rho
 R-E_0(\rho)}{\alpha}\nonumber\\
 &\geq&  \inf_{\alpha>0} \rho^* R + \frac{\rho^*
 R-E_0(\rho^*)}{\alpha}\nonumber\\
 &=& \rho^*R.\label{eqn:lowerbound_parameterization}
\end{eqnarray}

\begin{figure}[htbp]
\begin{center}
 \includegraphics[width=120mm]{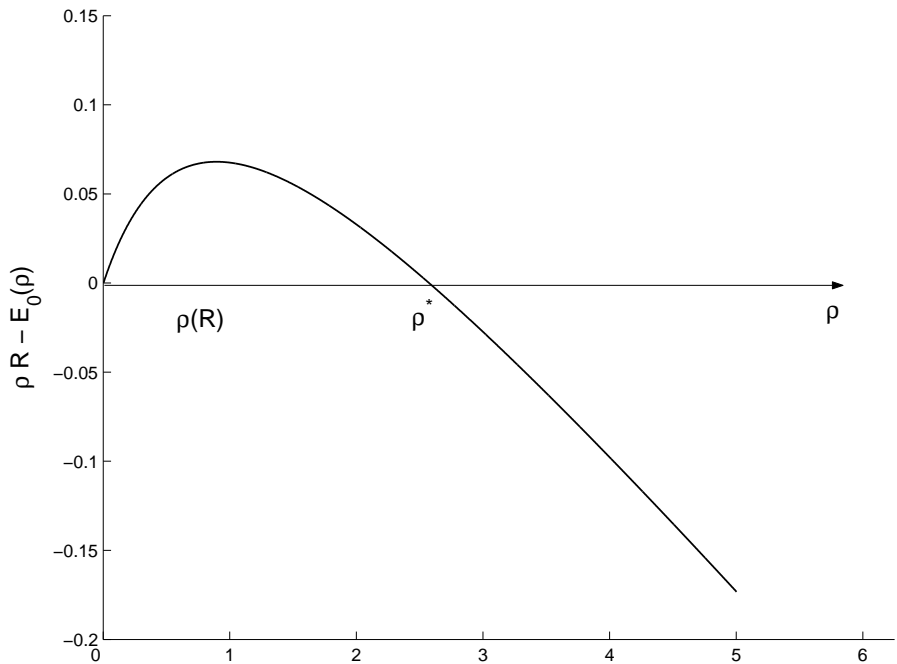}
\caption[]{Plot of $\rho R -E_0(\rho)$. The maximizing $\rho(R)<
  \rho^*$, the point where this crosses zero. $R$ is fixed.}
\label{fig.para_rho}
\end{center}
\end{figure}

Now show that $ E_{ei}(R) \leq \rho^* R$ by writing $\rho(R)$ as the
parameter $\rho$ that maximizes $\rho R -E_0(\rho)$.  From the
convexity of $E_0(\rho)$ for $\rho\in [0,\infty)$ and the fact that
$R\in [H(p_{\rvx|\rvy}),\log_2 M(p_{\rvx\rvy}))$, we know that
$\rho(R)$ is the unique positive real number s.t. $R=\frac{\partial
  E_0(\rho)}{\partial
  \rho}|_{\rho=\rho(R)}=H(\bar{p}^\rho_{\rvx|\rvy})|_{\rho=\rho(R)}$.
$\rho R -E_0(\rho)$ is a concave $\cap$ function of $\rho$ and $\rho R
-E_0(\rho)|_{\rho=0}=0$, hence $\rho(R)\leq \rho^*$ where $\rho(R)$ is
the maximal point and $\rho^*$ is the zero point of $\rho R
-E_0(\rho)$. This is illustrated in Figure~\ref{fig.para_rho}.

Because $R\in [H(p_{\rvx|\rvy}),\log_2 M(p_{\rvx\rvy}))$ and
$\rho(R)< \rho^*$, there exists $R'> R$, s.t. $\rho^*= \rho(R')$,
i.e. $\rho^*$ maximizes $\rho R'- E_0(\rho)$. Now let $\alpha^*=
\frac{R'}{R}-1$ which is positive because $R'>R$. That is $\rho^*$
maximizes $\rho R'- E_0(\rho) = \rho (1+\alpha^*)
 R-E_0(\rho )$. Plugging this in, gives:

\begin{eqnarray}
 E_{ei}(R)
 &=&\inf_{\alpha>0} \big\{\sup_{\rho\geq 0} \frac{\rho(\alpha+1)
 R-E_0(\rho)}{\alpha}\big\}\nonumber\\
 &\leq&  \sup_{\rho\geq 0}   \frac{\rho(1+\alpha^*)
 R-E_0(\rho)}{\alpha^*}\nonumber\\
 &=&    \frac{\rho^*(1+\alpha^*)
 R-E_0(\rho^*)}{\alpha^*}\nonumber\\
 &=& \rho^*R = E_0(\rho^*).\label{eqn:upperbound_parameterization}
\end{eqnarray}

Finally, to get the slope in the vicinity of the conditional entropy,
just expand $E_0(\rho)$ around $\rho=0$ using a Taylor series. The
constant term is zero and Lemma 17 of \cite{StreamingSlepianWolf}
reveals that the first order term is the conditional entropy
itself. The slope $\frac{\partial E}{\partial \rho} / \frac{\partial
  R}{\partial \rho}$ evaluated at $\rho = 0$ is clearly the
first-order term in the Taylor series divided by the second order
term, giving the desired result. The second derivative of $E_0(\rho)$
is only zero when $D(\bar{p}^0_{\rvx\rvy}\|p_{\rvx\rvy})=0$ which
implies that $p_{\rvx\rvy}$ is itself conditionally uniform, resulting
in the claimed infinite error exponents.

\section{Proof of Lemma~\ref{lemma.entropy-bound} } \label{app:lementropybound}

\begin{eqnarray}
nH(\rvx)&=_{(a)}&H(\rvx_1^n) \nonumber\\
&=& I(\rvx_1^n;\rvx_1^n)\nonumber\\
&=_{(b)}&I(\rvx_1^n; \rvxtil_1^n,\rvb_1^{\lfloor (n+\delay)R\rfloor},\rvy_1^{n+\delay})\nonumber\\
&=_{(c)}& I(\rvx_1^n;\rvy_1^{n+\delay})+  I(\rvx_1^n;\rvxtil_1^n|\rvy_1^{n+\delay}) + I(\rvx_1^n;\rvb_1^{\lfloor (n+\delay)R\rfloor}|\rvy_1^{n+\delay},\rvxtil_1^n )\nonumber\\
&\leq_{(d)} &  n I(\rvx ,\rvy )+ H(\rvxtil_1^n) + H(\rvb_1^{\lfloor (n+\delay)R\rfloor})\nonumber\\
&\leq & nH(\rvx)-nH(\rvx|\rvy) +H(\rvxtil_1^n)   + (n+\delay)R
\nonumber
\end{eqnarray}

$(a)$ is true because the source is iid. $(b)$ is true because of
the data processing inequality considering the following Markov
chain: $\rvx_1^n    \ \ - (\rvxtil_1^n,\rvb_1^{\lfloor
(n+\delay)R\rfloor},\rvy_1^n)  \ \ - \rvx_1^n$, thus
$I(\rvx_1^n;\rvx_1^n) \leq I(\rvx_1^n; \rvxtil_1^n,\rvb_1^{\lfloor
(n+\delay)R\rfloor},\rvy_1^{n+\delay})$. Furthermore,
$I(\rvx_1^n;\rvx_1^n)=H(\rvx_1^n)\geq I(\rvx_1^n; 
\rvxtil_1^n,\rvb_1^{\lfloor (n+\delay)R\rfloor},\rvy_1^{n+\delay})$.
Combining the two inequalities gives $(b)$.  $(c)$ is the chain rule
for mutual information. In $(d)$, first notice that $(\rvx,\rvy)$
are iid across time and thus
$I(\rvx_1^n;\rvy_1^{n+\delay})=I(\rvx_1^n;\rvy_1^{n})= n I(\rvx , 
\rvy)$. Second, the entropy of a random variable is never less than
the mutual information of that random variable with another one,
conditioned on another random variable or not. \hfill$\square$

\section{Proof of Lemma
  \ref{lemma.errorprobabilitybound}} \label{app:lemerrorprobbound} 

Lemma~\ref{lemma.entropy-bound} implies:
\begin{eqnarray}
\sum_{i=1}^n H(\rvxtil_i)&\geq& H(\rvxtil_1^n)\geq - (n+\delay)R +n
H(q_{\rvx|\rvy})
 \label{eqn.error_entropy_total}
\end{eqnarray}
The average entropy per source symbol for $\rvxtil$ is at least
$H(q_{\rvx|\rvy})- \frac{n+ \delay }{n}R $. Now suppose that
$H(\rvxtil_i)\geq \frac{1}{2}( H(q_{\rvx|\rvy}) -\frac{n+\Delta}{n}R)$
for $n_e$ symbol positions $1\leq j_1<j_2<... <j_A\leq n$. By noticing
that $H(\rvxtil_i)\leq \log_2|\mathcal{X}|$, we have
\begin{eqnarray}
\sum_{i=1}^n H(\rvxtil_i) \leq n_e \log_2 |\mathcal{X}| +
(n-n_e)\frac{1}{2}( H(q_{\rvx|\rvy}) -\frac{n+\Delta}{n}R)\nonumber
\end{eqnarray}
Combining this with (\ref{eqn.error_entropy_total}) gives:
\begin{eqnarray}
n_e \geq \frac{ (H(q_{\rvx|\rvy}) -\frac{n+\Delta}{n}R) }{
2\log_2|\mathcal{X}|-(H(q_{\rvx|\rvy}) -\frac{n+\Delta}{n}R)}n
\end{eqnarray}
where $2\log_2|\mathcal{X}|-(H(q_{\rvx|\rvy}) -\frac{n+\Delta}{n}R)
\geq 2\log_2|\mathcal{X}|-H(q_{\rvx|\rvy})\geq
2\log_2|\mathcal{X}|-\log_2|\mathcal{X}| >0$.

For each of the $j$, the individual entropy $H(\rvxtil_{j})\geq
\frac{1}{2}(H(q_{\rvx|\rvy})-\frac{n+ \delay }{n}R)$. By the
monotonicity of the binary entropy function, $\Pr(\rvxtil_{j}\neq x_0)
= \Pr(\rvx_{j}\neq \rvxhat_{j}) \geq \delta$.\hfill$\square$
\vspace{0.2in}

\section{Proof of
  Lemma~\ref{lemma.typicalset}} \label{app:lemtypicalset}

If we fix $\frac{\delay}{n}$ and let $n$ go to infinity, then by
definition $J = \min\{n, j^* + \delay\}$ goes to infinity as well. By
Lemma~13.6.1 in \cite{CoverThomas}, it is known that $\forall \epsilon
>0$, since $J- j^*$ and $j^*$ both getting large with $n$, that
$Q_{\rvx\rvy}(A^{\epsilon}_{J}(q_{\rvx\rvy})^C)\leq
\frac{\delta}{2}$. By Lemma~\ref{lemma.errorprobabilitybound},
$Q_{\rvx\rvy}(E_{j^*}) \geq \delta$. So
$$Q_{\rvx\rvy}(E_{j^*}\cap
A^{\epsilon}_{J}(q_{\rvx\rvy})) \geq Q_{\rvx\rvy}(E_{j^*} )
-Q_{\rvx\rvy}( A^{\epsilon}_{J}(q_{\rvx\rvy})^C) \geq
\frac{\delta}{2}$$ \hfill$\square$

\section{Proof of
  Lemma~\ref{lemma.ratiobound}} \label{app:lemratiobound}

For  $(\vec{x}, \svyBBar)\in
A^{\epsilon}_{J}(q_{\rvx\rvy})$, by the definition of the strongly typical
set, it can be easily shown by algebra that $
 D(r_{\bar{\svxBBar},\bar{\svyBBar}}\|p_{\rvx\rvy})\leq
D(q_{\rvx\rvy} \|p_{\rvx\rvy})+G\epsilon $ and  $
 D(r_{\xBar}\|p_{\rvx})\leq
D(q_{\rvx} \|p_{\rvx})+G\epsilon $. So

\begin{eqnarray}
\frac{p_{\rvx\rvy}(\vec{x}, \svyBBar)}{Q_{\rvx\rvy}(\vec{x},
\svyBBar)}&=&\frac{p_{\rvx\rvy}(\xBar)}{q_{\rvx\rvy}(\xBar)}\frac{p_{\rvx\rvy}(\bar{\svxBBar},
\bar{\svxBBar})}{q_{\rvx\rvy}(\svxBBar, \svyBBar)}
\frac{p_{\rvx\rvy}(\svx_{J+1}^{j^*+\delay}, \svy_{J+1}^{j^*+\delay})}{p_{\rvx\rvy}(\svx_{J+1}^{j^*+\delay}, \svy_{J+1}^{j^*+\delay})}\nonumber\\
&=&\frac{2^{-(J-j^*+1)(D(r_{\bar{\svxBBar},\bar{\svyBBar}}\|p_{\rvx\rvy})+H(r_{\bar{\svxBBar},\bar{\svyBBar}}))
}}{2^{-(J-j^*+1)(D(r_{\bar{\svxBBar},\bar{\svyBBar}}\|q_{\rvx\rvy})+H(r_{\bar{\svxBBar},\bar{\svyBBar}}))
}} \frac{2^{-(j^*-1)(D(r_{\xBar}\|p_{\rvx})+H(r_{\xBar})) }}{2^{-(j^*-1)(D(r_{\xBar}\|q_{\rvx})+H(r_{\xBar })) }}\nonumber\\
&\geq_{(a)}& 2^{-(J-j^*+1) (D(q_{\rvx\rvy}\|p_{\rvx\rvy})+G\epsilon
)
-(j^*-1) (D(q_{\rvx}\|p_{\rvx})+G \epsilon )}  \nonumber\\
& =_{}&  2^{-(J-j^*+1)  D(q_{\rvx\rvy}\|p_{\rvx\rvy}) -(j^*-1)
D(q_{\rvx}\|p_{\rvx}) -JG \epsilon}\nonumber
\end{eqnarray}
where $(a)$ is true by (12.60) in \cite{CoverThomas}. \hfill $\square$

\section{Proof of
  Lemma~\ref{lemma.proofoftheorem}} \label{app:lemproofoftheorem}

Combining Lemmas \ref{lemma.typicalset} and
\ref{lemma.ratiobound}:
\begin{eqnarray}
 p_{\rvx\rvy}(E_{j^*}) &\geq & p_{\rvx\rvy}(E_{j^*}\cap
A^{\epsilon}_{J}(q_{\rvx\rvy}))
\nonumber\\
& \geq & q_{\rvx\rvy}(E_{j^*}\cap
A^{\epsilon}_{J}(q_{\rvx\rvy}))2^{-(J-j^*+1)
D(q_{\rvx\rvy}\|p_{\rvx\rvy}) -(j^*-1)
D(q_{\rvx}\|p_{\rvx}) -JG \epsilon}\nonumber\\
& \geq& \frac{\delta}{2}2^{-(J-j^*+1)  D(q_{\rvx\rvy}\|p_{\rvx\rvy})
-(j^*-1) D(q_{\rvx}\|p_{\rvx}) -JG \epsilon}
   \nonumber
\end{eqnarray}
\hfill $\square$

\section{Proof of
  Corollary~\ref{corollary_SI_upper_symmetric}} \label{app:corsisymmetric}
Theorem~\ref{THM_UPPER_SI} asserts that
\begin{eqnarray}
E_{si}^{ }(R) &\leq &  \{\inf_{q_{\rvx\rvy}, \alpha\geq 1:
H(q_{\rvx|\rvy})>(1+\alpha)R
}\{\frac{1}{\alpha}D(q_{\rvx\rvy}\|p_{\rvx\rvy}) \}, \nonumber\\
&&\inf_{q_{\rvx\rvy}, 1\geq \alpha\geq 0:
H(q_{\rvx|\rvy})>(1+\alpha)R
}\{\frac{1-\alpha}{\alpha}D(q_{\rvx}\|p_{\rvx})
+D(q_{\rvx\rvy}\|p_{\rvx\rvy}) \} \}\nonumber\\
&\leq & \inf_{q_{\rvx\rvy}, 1\geq \alpha\geq 0:
H(q_{\rvx|\rvy})>(1+\alpha)R
}\{\frac{1-\alpha}{\alpha}D(q_{\rvx}\|p_{\rvx})
+D(q_{\rvx\rvy}\|p_{\rvx\rvy}) \}  \nonumber\\
&\leq & \inf_{q_{\rvx\rvy}, 1\geq \alpha\geq 0:
H(q_{\rvx|\rvy})>(1+\alpha)R, \ \ q_\rvx=p_\rvx
}\{\frac{1-\alpha}{\alpha}D(q_{\rvx}\|p_{\rvx})
+D(q_{\rvx\rvy}\|p_{\rvx\rvy}) \}  \nonumber\\
&=& \inf_{q_{\rvx\rvy}, 1\geq \alpha\geq 0:
H(q_{\rvx|\rvy})>(1+\alpha)R, \ \ q_\rvx=p_\rvx }\{  D(q_{\rvx\rvy}\|p_{\rvx\rvy}) \}  \nonumber\\
&=& \inf_{q_{\rvx\rvy}: H(q_{\rvx|\rvy})>R, \ \ q_\rvx=p_\rvx }\{
D(q_{\rvx\rvy}\|p_{\rvx\rvy}) \} \label{eqn:uniform_SI_upper}
\end{eqnarray}

The next step is to show that (\ref{eqn:uniform_SI_upper}) is indeed
$E_{s,b}(R, p_\rvs)$ for uniform sources $\rvx$ and symmetric side
information $\rvy$, where $\rvx=\rvy\oplus\rvs$.

\begin{eqnarray} \inf_{q_{\rvx\rvy}:
H(q_{\rvx|\rvy})>R, \ \ q_\rvx=p_\rvx}\{
D(q_{\rvx\rvy}\|p_{\rvx\rvy}) \} &=_{(a)}& \inf_{q_{\rvx\rvy}:
H(q_{\rvx|\rvy})>R}\{ D(q_{\rvx\rvy}\|p_{\rvx\rvy}) \}\label{eqn:corollary_proof_symmetricsource}\\
&=_{(b)} & \max_{\rho\geq 0} \rho R- E_0(\rho)\nonumber\\
&=_{(c)}&\max_{\rho\geq 0} \rho R- (1+\rho)\log [ \sum_s
 p_{\rvs}(s)^{\frac{1}{1+\rho}}] \nonumber\\
&=_{(d)}& E_{s,b} (R, p_\rvs)\nonumber
\end{eqnarray}
$(b)$ follows from
(\ref{eqn:blockupperboundrho}) in Theorem~\ref{THM.SI_BLOCK}, $(c)$
follows since (\ref{eqn:gallagerfunction}) can be simplified for this case:
\begin{eqnarray}
 E_0(\rho) & = &
 \log_2\sum_{y}(\sum_{x} p_{\rvx\rvy}(x,y)^{\frac{1}{1+\rho}})^{(1+\rho)} \nonumber\\
 & = &
 \log_2\sum_{y}(\sum_{x} (p_{\rvy}(y)p_{\rvx|\rvy}(x|y))^{\frac{1}{1+\rho}})^{(1+\rho)} \nonumber\\
 & = &
 \log_2\sum_{y} p_{\rvy}(y) (\sum_{x}
 (p_{\rvx|\rvy}(x|y))^{\frac{1}{1+\rho}})^{(1+\rho)} \nonumber\\
 & = &
 \log_2\sum_{y} p_{\rvy}(y) (\sum_{s}
 (p_{\rvs}(s)^{\frac{1}{1+\rho}})^{(1+\rho)} \nonumber\\
 & = &
 \log_2 (\sum_{s}
 (p_{\rvs}(s)^{\frac{1}{1+\rho}})^{(1+\rho)}\nonumber\\
 & = & (1+\rho) \log_2 (\sum_{s}p_{\rvs}(s)^{\frac{1}{1+\rho}})\label{eqn:simplify_E0}
\end{eqnarray}
where this clearly matches from
(\ref{eqn:gallagerfunction_singlesource}) to give us $(d)$. 

Thus, for uniform source $\rvx$ and side information
$\rvy=\rvx\ominus \rvs$, the distribution $q_{\rvx\rvy}$ that
minimizes the RHS of (\ref{eqn:corollary_proof_symmetricsource}) is
also marginally uniform on $\rvx$ since all that needs to tilt is the
distribution for $\rvs$. Hence the constraint on the
marginal $q_\rvx=p_\rvx$ is redundant and $(a)$ is true.  \hfill $\square$

\bibliography{IEEEabrv,MyMainBibliography}
\bibliographystyle{IEEEtran}

\end{document}